\documentclass[final,5p,times,twocolumn]{elsarticle}

\usepackage{amssymb}
\usepackage{amsthm}
\newdefinition{assumption}{Assumption}

\usepackage{amsmath}
\usepackage{siunitx}
\usepackage{ulem}
\usepackage{natbib}
\usepackage{color}					
\usepackage{soul}					

\usepackage{nomencl}				
\makenomenclature
\usepackage{ifthen}
\renewcommand{\nomgroup}[1]{%
	\ifthenelse{\equal{#1}{A}}{\item[\textbf{Latin symbols}]}{%
		\ifthenelse{\equal{#1}{B}}{\item[\textbf{Greek symbols}]}{%
			\ifthenelse{\equal{#1}{C}}{\item[\textbf{Subscripts and Superscripts}]}{}}}%
}

\newcommand{\noml}[1][]{\nomenclature[A#1]} 	
\newcommand{\nomg}[1][]{\nomenclature[B#1]} 	
\newcommand{\noms}[1][]{\nomenclature[C#1]} 	

\usepackage{tikz}
\usetikzlibrary{arrows}
\usetikzlibrary{patterns}
\usetikzlibrary{shapes.arrows}
\usepackage{calc}
\newlength{\halfdiffT}
\newlength{\halfdiffQ}
\newlength{\halfdiffTset}
\newlength{\halfdiffTamb}
\usepackage[final]{draftwatermark}
\SetWatermarkText{Draft}
\SetWatermarkScale{2}

\journal{Control Engineering Practice}

\begin{document}

\begin{frontmatter}



\title{Dynamic state-space modeling and model-based control design for loop heat pipes}


\author[label1]{Thomas Gellrich}
\author[label1]{Clara Sester}
\author[label2]{Max Okraschevski}
\author[label1]{Stefan Schwab}
\author[label2]{Hans-Joerg Bauer}
\author[label3]{Soeren Hohmann}

\address[label1]{FZI Research Center for Information Technology, Karlsruhe, Germany, gellrich@fzi.de}
\address[label2]{Institute of Thermal Turbomachinery, Karlsruhe Institute of Technology, Karlsruhe, Germany, max.okraschevski@kit.edu}
\address[label3]{Institute of Control Systems, Karlsruhe Institute of Technology, Karlsruhe, Germany, soeren.hohmann@kit.edu}

\begin{abstract}
For the thermal control of electronic components in aerospace, automotive or server systems, the heat sink is often located far from the heat sources. Therefore, heat transport systems are necessary to cool the electronic components effectively. Loop heat pipes (LHPs) are such heat transport systems, which use evaporation and condensation to reach a higher heat transfer coefficient than with sole heat conduction. The operating temperature of the LHP governs the temperature of the electronic components, but depends on the amount of dissipated heat and the temperature of the heat sink. For this reason, a control heater on the LHP provides the ability to control the operating temperature at a fixed setpoint temperature. For the model-based control design of the control heater controller, the current LHP state-space model in the literature focuses on the setpoint response without modeling the fluid's dynamics. However, the fluid's dynamics determine the disturbance behavior of the LHP. Therefore, the fluid's dynamics are incorporated into a new LHP state-space model, which is not only able to simulate the LHP behavior under disturbance changes, but is also used for the model-based design of a robust nonlinear controller, which achieves an improved control performance compared to the nonlinear controller based on the previous LHP state-space model.
\end{abstract}

%


\end{frontmatter}



%

\noml[A]{$A_c$}{Cross-sectional area of transport line/condenser, \si{\square\meter}}
\noml[Aa]{$A_{wf}$}{Substance specific parameter in Antoine equation}
\noml[B]{$B_{wf}$}{Substance specific parameter in Antoine equation}
\noml[C]{$C_{wf}$}{Substance specific parameter in Antoine equation}
\noml[cp]{$c_p$}{Specific heat capacity, \si{\joule \per \kilogram \per \kelvin} }
\noml[D]{$D_c$}{Diameter of transport line/condenser, \si{\meter} }
\noml[hv]{$\Delta h_v$}{Specific heat of evaporation, \si{\joule \per \kilogram} }
\noml[h]{$h$}{Specific enthalpy, \si{\joule \per \kilogram} }
\noml[Rb]{$R$}{Specific gas constant, \si{\joule \per \kilogram \per \kelvin} }
\noml[Rw]{$R_{wick}$}{Thermal resistance of wick, \si{\kelvin \per \watt} }
\noml[Rp]{$R_p$}{Pore radius, \si{\meter} }
\noml[m]{$m$}{Mass, \si{\kilogram} }
\noml[mdot]{$\dot{m}$}{Mass flow rate, \si{\kilogram \per \second}}
\noml[temp]{$T$}{Temperature, \si{\kelvin} or \si{\celsius} }
\noml[time]{$t$}{Time in continuous form, \si{\second} }
\noml[V]{$V$}{Volume, \si{\cubic\meter} }
\noml[Q]{$\dot Q$}{Rate of heat flow, \si{\watt} }
\noml[k]{$k$}{Heat transfer coefficient, \si{\joule\per\kilogram} }
\noml[x]{$x$}{Vapor quality}
\noml[p]{$p$}{Pressure, \si{\pascal} }
\noml[L]{$L$}{Length, \si{\meter} }

\nomg[a]{$\bar{\alpha}$}{Mean void fraction}
\nomg[e]{$\eta$}{Length of two-phase region in condenser, \si{\meter} }
\nomg[m]{$\mu$}{Dynamic viscosity, \si{\pascal \second}}
\nomg[theta]{$\theta$}{Contact angle, \si{\degree}}
\nomg[sigma]{$\sigma$}{Surface tension, \si{\newton \per \meter}}
\nomg[rho]{$\rho$}{Density, \si{\kilogram \per \cubic\meter}}

\noms[cc]{cc}{Compensation chamber}
\noms[l]{l}{Liquid}
\noms[v]{v}{Vapor}
\noms[wick]{wick}{Wick}
\noms[ccin]{cc,in}{Inlet of compensation chamber}
\noms[evap]{evap}{Evaporator}
\noms[amb]{amb}{Ambient}
\noms[wf]{wf}{Working fluid}
\noms[cap]{cap}{Capillary}
\noms[p]{p}{Pore}
\noms[c]{c}{Cross-section of transport line/condenser}
\noms[cond]{cond}{Condenser}
\noms[LL]{ll}{Liquid line}
\noms[z]{2$\varphi$}{Two-phase region of condenser}
\noms[sink]{sink}{Sink}
\noms[sc]{sc}{Subcooled region of condenser}
\noms[fri]{fri}{Friction}
\noms[op]{op}{Operating point}
\noml[zzz]{}{}   
\nomg[zzz]{}{}


\printnomenclature


\section{Introduction}
\label{sec:intro}
Loop heat pipes (LHPs) are passive, two-phase heat transport systems with their origin in the thermal control of aerospace systems \citep{Maydanik2005}. In recent years, their application has spread to other technology areas such as server systems (\cite{Zimbeck2008}, \cite{Maydanik2010}) and automotive systems (\cite{Putra2016}, \cite{Bernagozzi2018}). Their advantage is based on an effective and reliable heat transport with a working fluid between distributed heat sources and heat sinks. The phase of the working fluid inside the LHP is changed through evaporation and condensation to achieve a higher heat transfer coefficient than with direct heat conduction. Furthermore, their passive working principle enables a mass flow in the loop between the locally separated evaporator and condenser over long distances compared to traditional heat pipes by using capillary forces in a porous wick inside the evaporator instead of an error-prone mechanical pump. However, the operating temperature of the LHP varies with the operating conditions during heat transportation and must be controlled with an additional control heater in order to stay in a desired temperature corridor (\cite{Ku2008}). The changes of the operating conditions are traced back to mainly two causes: On the one hand, the sink temperature varies due to external influences on the heat sink, e.g. fluctuating insolation on the radiator in the orbit or variable air movements at the air cooler on earth. On the other hand, the electronic components dissipate different amounts of excess heat depending on their operating statuses that form the total heat load at the evaporator.

Simple control algorithms for the control heater are used experimentally to control the LHP operating temperature (heuristic PID controllers (\cite{Ku2011}, \cite{Ku2011a}) and two-point controllers (\cite{Khrustalev2014}, \cite{Ku2011})). However, model-based controllers for control heaters to overcome the complex performance characteristics of LHPs are rare in the literature owing to the fact that most LHP models (e.g. \cite{Riehl2005}, \cite{Kaya2008}, \cite{Chernysheva2008}, \cite{Bernagozzi2018}, \cite{Meinicke2019}) are of numerical nature for LHP design characterization and LHP simulation, but unsuitable for model-based control design. Similar focus on the LHP design is set with analytical LHP models in \cite{Shukla2008} and \cite{Li2010}. In \cite{Launay2007}, differential equations for the dynamic behavior of the LHP without a control heater are established to analyze the LHP temperature oscillations at the borders of the LHP operating range without presenting the solution procedure of the system of equations.

A first model-based control design approach is motivated in \cite{Gellrich2018a} by the proportional time-delaying behavior of the CC temperature, which controls the operating temperature of the LHP \citep{Ku1999}. The transfer function of the CC temperature as linear black box model is identified to calculate the PI parameters of a two-degree-of-freedom controller for improved disturbance response. In \cite{Gellrich2018b}, the PI parameters are physically motivated with a linearized LHP state-space model and the stationary mass flow rate as system parameter. However, to improve the robustness of the LHP operating temperature control against the aforementioned influences, more sophisticated control algorithms for the control heater are necessary \citep{Ku2011a}. In \cite{Gellrich2019}, a nonlinear model identification adaptive control is designed by interpreting the underlying LHP model of \cite{Gellrich2018b} as nonlinear parameter-varying system. Introducing an online parameter estimation and temperature prediction, the time variance of the transient mass flow rate is included to achieve an improved performance of the operating temperature control without extending the underlying control model.

In this work, an overall LHP state-space model is derived that physically describes the dynamics of the temperatures and the mass flows in consideration of variable disturbances.  By tracking the vapor-liquid interface in the condenser and the liquid volume in the compensation chamber, the analytical model is able to predict the transient behavior of the LHP over the entire operating range. A nonlinear controller is designed based on the introduced model that reacts to disturbances without the necessity of a numerical parameter estimation because of the underlying dynamic description of the mass flows. 

This paper is structured as follows: In Section~2, the fundamentals of the LHP operating characteristics are presented. After the state-space modeling of the LHP in Section~3, the model-based nonlinear controller is designed in Section~4. The performances of the LHP model and the nonlinear controller are evaluated in Section~5, followed by the conclusions in Section~6.

\section{LHP characteristics}
\label{sec:fund}
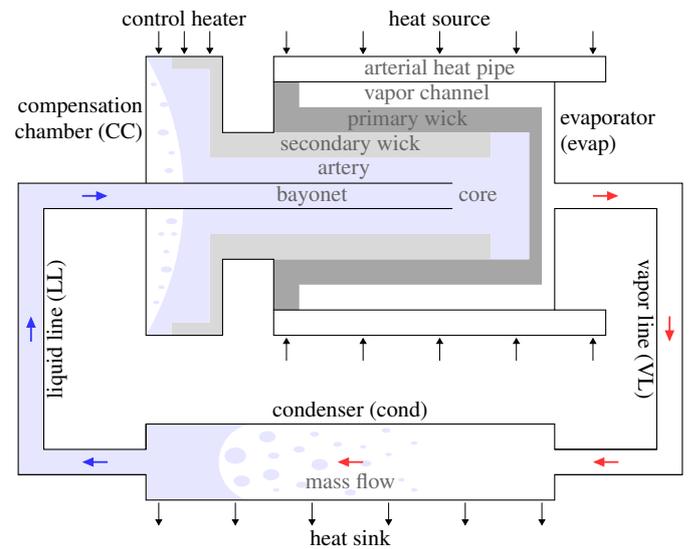
\begin{figure}[htb]
	\centering
	\resizebox{\linewidth}{!}{\begin{tikzpicture}
	\usetikzlibrary{patterns}
	
	\tikzstyle{every node}=[font={\fontsize{25}{0}\selectfont}]
	\tikzstyle{->} = [thick, -triangle 60];
	
	\definecolor{redish}{RGB}{255,50,50}
	\definecolor{bluish}{RGB}{50,50,255}
	\definecolor{red}{RGB}{255,0,0}
	\definecolor{green}{RGB}{0,153,0}
	\definecolor{violet}{RGB}{128,0,128}
	\definecolor{greenish}{RGB}{0,176,80}
	
	\draw[blue!10,fill] (3.75,-11.25) ellipse (3mm and 2mm);
	\draw[blue!10,fill] (4.75,-9.5) ellipse (3mm and 2mm);
	\draw[blue!10,fill] (4.9,-11) ellipse (3mm and 2mm);
	\draw[blue!10,fill] (3.5,-10.25) ellipse (4mm and 3mm);
	\draw[blue!10,fill] (6,-10) ellipse (2mm and 1mm);
	\draw[blue!10,fill] (5,-10.25) ellipse (2mm and 1mm);
	\draw[blue!10,fill] (6.1,-11.4) ellipse (2mm and 1mm);
	\draw[blue!10,fill] (8,-11.7) ellipse (2mm and 1mm);
	\draw[blue!10,fill] (10,-11.2) ellipse (1mm and 0.5mm);
	\draw[blue!10,fill] (9.4,-9.25) ellipse (1.5mm and 0.75mm);
	\draw[blue!10,fill] (6.8,-10.6) ellipse (3mm and 2mm);
	\draw[blue!10,fill] (10.5,-9.75) ellipse (1mm and 0.5mm);
	\draw[blue!10,fill] (8.5,-9.9) ellipse (1mm and 0.5mm);
	\draw[blue!10,fill] (7.5,-9.8) ellipse (1.5mm and 0.75mm);
	\draw[blue!10,fill] (9.5,-10.2) ellipse (1mm and 0.5mm);
	\draw[blue!10,fill] (6.6,-9.4) ellipse (1.5mm and 0.75mm);
	\draw[blue!10,fill] (5.1,-11.6) ellipse (1mm and 0.5mm);
	\draw[blue!10,fill] (10.5,-11.6) ellipse (1mm and 0.5mm);
	\draw[blue!10,fill] (8.6,-10.55) ellipse (2mm and 1mm);
	
	\fill[black!35] (15,2.5) -- (5,2.5) -- (5,4.5) -- (6,4.5) -- (6,3.5) -- (15.5,3.5) -- (15.5,-3.5) -- (6,-3.5) -- (6,-4.5) -- (5,-4.5) -- (5,-2.5) -- (15,-2.5);
	\fill[black!15] (13.5,-1.5) -- (2.5,-1.5) -- (2.5,-5) -- (1,-5) -- (1,-5.5) -- (3,-5.5) -- (3,-2.5) -- (13.5,-2.5);
	\fill[black!15] (13.5,2.5) -- (3,2.5) -- (3,5.5) -- (1,5.5) -- (1,5) -- (2.5,5) -- (2.5,1.5) -- (13.5,1.5);
	\fill[blue!10] (0,-9) -- (0,-10) -- (-4,-10) -- (-4,-0.5) -- (0,-0.5) -- (12,-0.5)  arc (75:-75:-0.52) -- (0,0.5) -- (-5,0.5) -- (-5,-11) -- (0,-11) -- (0,-12) -- (4,-12) arc (75:-75:-1.55) ;
	\fill[blue!10] (0,5.5) arc (30:-30:11) -- (0,-5.5) -- (1,-5.5) -- (1,-5) -- (2.5,-5) -- (2.5,-1.5) -- (13.5,-1.5) -- (13.5,-2.5) -- (15,-2.5) -- (15,2.5) -- (13.5,2.5) -- (13.5,1.5) -- (2.5,1.5) -- (2.5,5) -- (1,5) -- (1,5.5);
 	\draw[blue!10,fill] (0.9,1) ellipse (1.5mm and 0.75mm);
	\draw[blue!10,fill] (1,1.7) ellipse (1mm and 0.5mm);
	\draw[blue!10,fill] (0.8,-2.2) ellipse (1mm and 0.5mm);
	\draw[blue!10,fill] (1,-1) ellipse (2mm and 1mm);
	\draw[blue!10,fill] (0.6,-3.6) ellipse (1mm and 0.5mm);
	\draw[blue!10,fill] (0.7,3.6) ellipse (1.5mm and 0.75mm);
	\draw[blue!10,fill] (0.6,-1.5) ellipse (1mm and 0.5mm);
	\draw[blue!10,fill] (0.675,2.5) ellipse (2mm and 1mm);
	\draw[blue!10,fill] (0.8,-2.9) ellipse (2mm and 1mm);
	\draw[blue!10,fill] (0.3,4.4) ellipse (1mm and 0.5mm);
	\draw[blue!10,fill] (1.1,-1.5) ellipse (1mm and 0.5mm);
	\draw[blue!10,fill] (0.4,-4.2) ellipse (1mm and 0.5mm);
 	
	\draw[very thick] (0,0.5) -- (0,5.5) -- (3,5.5) -- (3,2.5) -- (5,2.5) -- (5,4.5) -- (16,4.5) -- (16,0.5) -- (21,0.5) -- (21,-11) -- (16,-11);
	\draw[very thick] (0,-0.5) -- (0,-5.5) -- (3,-5.5) -- (3,-2.5) -- (5,-2.5) -- (5,-4.5) -- (16,-4.5) -- (16,-0.5) -- (20,-0.5) -- (20,-10) -- (16,-10);
	\draw[very thick] (0,-9) -- (0,-10) -- (-4,-10) -- (-4,-0.5) -- (0,-0.5) -- (12,-0.5);
	\draw[very thick] (12,0.5) -- (0,0.5) -- (-5,0.5) -- (-5,-11) -- (0,-11);
	\draw[very thick] (0,-10) -- (0,-9) -- (16,-9) -- (16,-10) -- (20,-10);
	\draw[very thick] (-5,-11) -- (0,-11) -- (0,-12) -- (16,-12) -- (16,-11) -- (21,-11);
	
	
	\node[black!60] at (11.5,5) {arterial heat pipe}; 
	\draw[very thick] (5,4.5) -- (5,5.5) -- (18,5.5) -- (18,4.5) -- (16,4.5);
	\draw[very thick] (5,-4.5) -- (5,-5.5) -- (18,-5.5) -- (18,-4.5) -- (16,-4.5);
	\node[] at (11.5,7) {heat source}; 
	\draw[->] (5.5,6.5) -- (5.5,5.6);
	\draw[->] (8.5,6.5) -- (8.5,5.6);
	\draw[->] (11.5,6.5) -- (11.5,5.6);
	\draw[->] (14.5,6.5) -- (14.5,5.6);
	\draw[->] (17.5,6.5) -- (17.5,5.6);
	\draw[->] (5.5,-6.5) -- (5.5,-5.6);
	\draw[->] (8.5,-6.5) -- (8.5,-5.6);
	\draw[->] (11.5,-6.5) -- (11.5,-5.6);
	\draw[->] (14.5,-6.5) -- (14.5,-5.6);
	\draw[->] (17.5,-6.5) -- (17.5,-5.6);
	
%
	
	\node[] at (1.5,7) {control heater}; 
	\draw[->] (0.5,6.5) -- (0.5,5.6);
	\draw[->] (1.5,6.5) -- (1.5,5.6);
	\draw[->] (2.5,6.5) -- (2.5,5.6);
	
	\node[] at (8,-13.5) {heat sink}; 
	\draw[->] (0.5,-12.1) -- (0.5,-13);
	\draw[->] (3.5,-12.1) -- (3.5,-13);
	\draw[->] (6.5,-12.1) -- (6.5,-13);
	\draw[->] (9.5,-12.1) -- (9.5,-13);
	\draw[->] (12.5,-12.1) -- (12.5,-13);
	\draw[->] (15.5,-12.1) -- (15.5,-13);
	
	\node[black!60] at (8,-11.25) {mass flow};	
	\draw[-triangle 45, line width=2pt, redish] (17.5,0) -- (18.5,0); 
	\draw[-triangle 45, line width=2pt, redish] (20.5,-4.75) -- (20.5,-5.75); 
	\draw[-triangle 45, line width=2pt, redish] (18.5,-10.5) -- (17.5,-10.5); 
	\draw[-triangle 45, line width=2pt, redish] (8.5,-10.5) -- (7.5,-10.5); 
	\draw[-triangle 45, line width=2pt, bluish] (-1.5,-10.5) -- (-2.5,-10.5);
	\draw[-triangle 45, line width=2pt, bluish] (-4.5,-5.75) -- (-4.5,-4.75);	
	\draw[-triangle 45, line width=2pt, bluish] (-2.5,0) -- (-1.5,0);	
	
	\node[black!60] at (6.5,0) {bayonet}; 
	\node[black!60] at (10.25,3) {primary wick}; 
	\node[black!60] at (8,2) {secondary wick}; 
	\node[black!60] at (13,0) {core}; 
	\node[black!60] at (11,4) {vapor channel}; 
	\node[black!60] at (7.75,1) {artery}; 

	\node[black,above] at (8,-9) {condenser (cond)}; 
	\node[rotate=90,below] at (-4,-5.25) {liquid line (LL)}; 
	\node[rotate=-90,below] at (20,-5.25) {vapor line (VL)}; 
	\node[black,left] at (0,3.5) {compensation}; 
	\node[black,left] at (0,2.5) {chamber (CC)}; 
	\node[black,right] at (16.1,3.0) {evaporator}; 
	\node[black,right] at (16.1,2.0) {(evap)}; 
	
\end{tikzpicture}}
	\caption{Schematic of the LHP (cf. \cite{Ku1999})}
	\label{fig:lhp}
\end{figure}

The schematic structure of the LHP is shown in Fig.~\ref{fig:lhp}.

The LHP consists of five components: the compensation chamber (CC), the evaporator, the condenser, the liquid line (LL), and the vapor line (VL). Inside the evaporator, the primary wick is a porous metal structure, which builds up capillary forces when the working fluid evaporates at the wick's outlet due to the heat load transported to the LHP by arterial heat pipes. These capillary forces induce the circulation of the working fluid in the loop. The vaporized working fluid flows from the evaporator through the VL to the condenser, where it condenses by transferring its heat to the heat sink. The temperature of the remaining liquid in the condenser is further decreased through subcooling. By the pressure difference built up in the primary wick, the working fluid is pushed onward from the condenser through the LL to the CC. The CC stores excess fluid and conduces to the control of the LHP operating temperature. Since a dryout of the primary wick disrupts the fluid circulation and thus the continuous operation of the LHP, a secondary wick is incorporated into the CC to ensure a sufficient supply of liquid to the primary wick at all times \citep{Ku1999}.
\begin{figure}[htb]
	\centering
	\resizebox{1.00\linewidth}{!}{\begin{tikzpicture}	
	\tikzstyle{every node}=[font={\huge}]
	\tikzstyle{->} = [very thick, -triangle 60];
	\draw[->] (-0.2,0.5) -- (14,0.5) node[right] {$\dot{Q}$};
	\draw[->] (0,0.3) -- (0,14) node[above] {$T$};
	\draw[color=black] (-0.2,9.5) node[left]{$T_{set}$} -- (0.2,9.5);
	\draw[color=black] (-0.2,2) node[left]{$T_{amb}$} -- (0.2,2);
	\draw[color=black] (1.6,0.3) node[below]{$\dot{Q}_{low}$} -- (1.6,0.7);
	\draw[color=black,dashed] (1.6,0.8) -- (1.6,9.5);
	\draw[color=black] (12,0.3) node[below]{$\dot{Q}_{high}$} -- (12,0.7);
	\draw[color=black,dashed] (12,0.8) -- (12,2.65);
	\draw[color=black,dashed] (12,5.25) -- (12,9.5);
	\filldraw[color=red!60,opacity=0.5] (1.6,9.5) -- (1.8,9)..controls(4,3.5)..(11.8,9.35) -- (12,9.5);
	\filldraw[color=green!60!black,opacity=0.5] (12,9.5) -- (13.9,10.925) -- (13.9,9.5);
	\filldraw[color=cyan!60,opacity=0.5] (0.6,12) -- (1.6,9.5) -- (0.6,9.5);
	\draw[color=black!75!white] (10,0.5) -- (10,12.25);
	\draw[color=black] (10,0.3) node[below]{$\dot{Q}_{trans}$} -- (10,0.7);
	\node[above,right,align=left,color=black!75!white] at (10.25,4) {fixed\\conductance\\mode};
	\node[above,left,align=right,color=black!75!white] at (9.75,4) {variable\\conductance\\mode};
	\draw[color=black] (-0.2,0.5) node[left]{$T_{sink}$} -- (0.2,0.5);
	\draw[color=black] (-0.2,4.92) node[left]{$T_{ot,min}$} -- (0.2,4.92);
	\draw[color=black,dashed] (0.3,4.92)-- (4.63,4.92);
	\draw[color=black] (4.62,0.3) node[below]{$\dot{Q}_{ot,min}$} -- (4.62,0.7);
	\draw[color=black,dashed] (4.62,4.92)-- (4.62,0.8);
	\draw[color=black, very thick] (0.5,12.25) -- (1.8,9)..controls(4,3.5)..(11.8,9.35)  -- node[above,at end,sloped,xshift=-200,yshift=4]{natural SSOT} (14,11); 
	\draw[color=blue, very thick] (0,9.5) -- node[above,sloped]{fixed SSOT} (14,9.5);
	\draw[color=black,domain=0:11,dashed] plot (\x,0.75*\x+0.5) {};
	\node[above,color=black!50!white] at (7.5,0.5) {heat load};
	\node[below,rotate=90,color=black!50!white] at (0,7.25) {temperature};
	\node[color=red,align=center] at (6,13.25) {CC\\heating};
	\node[color=green!60!black,align=center] at (12.75,13.25) {condenser\\increasing};
	\node[right,color=cyan,align=center] at (0,13.25) {CC\\cooling};

\end{tikzpicture}}
	\caption{Steady-state operating temperature (SSOT) of an LHP (cf. \cite{Ku2008} and \cite{Chuang2003})}	
	\label{Abb:op_charac}
\end{figure}
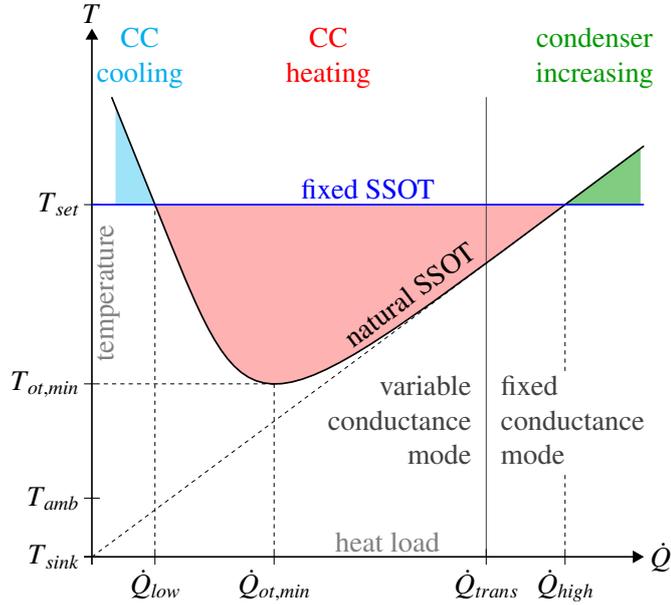

Fig.~\ref{Abb:op_charac} shows the steady-state operating characteristics of a LHP. The graph describes the steady-state operating temperature (SSOT) as a function of the heat load. When the ambient temperature $T_{amb}$ is higher than the sink temperature $T_{sink}$, the typical U-shaped curve of the natural SSOT is generated by the influence of the heat input from the surroundings into the LL (cf. \cite{Chuang2003}). At low heat loads, little vapor is generated in the evaporator inducing low mass flow rates in the VL and the LL. While passing the LL, the liquid is heated by the surroundings. The higher the heat load, the higher the mass flow rates and the shorter the residence time of the liquid in the LL to absorb heat from the surroundings. As a consequence, more subcooling arrives at the CC to compensate the heat leak from the evaporator and the SSOT decreases \citep{Ku1999}. At the same time, the higher the heat load, the more the liquid-vapor interface in the condenser proceeds in the direction of the condenser outlet. Thus, the liquid in the condenser is not fully subcooled to sink temperature anymore and the SSOT rises. At the heat load $\dot{Q}_{ot,min}$, the lowest SSOT is reached as a combination of high mass flow rates and a low condenser outlet temperature. When the heat load is higher than $\dot{Q}_{trans}$, the LHP operation mode changes from \textit{variable conductance mode} to \textit{fixed conductance mode}. In \textit{variable conductance mode}, not only condensing but also subcooling of the condensate takes place in the condenser. This means that the condenser is not yet fully utilized and the LHP has not yet reached its maximum heat transfer capability. In \textit{fixed conductance mode}, the vapor-liquid interface is located at the condenser outlet leaving no space for subcooling. The LHP operates with its maximum overall heat transfer coefficient and the SSOT rises linearly with increasing heat load \citep{Chuang2003}.

In order to control the natural SSOT at a fixed SSOT, several actions are required depending on the amount of heat load applied to the evaporator. For heat loads between $\dot{Q}_{low}$ and $\dot{Q}_{high}$, the CC is heated to reach the fixed SSOT (red area), while for heat loads lower than $\dot{Q}_{low}$, the CC is cooled (blue area). For heat loads higher than $\dot{Q}_{high}$, the condenser needs to be enlarged constructively (green area) to loose the heat to the heat sink while preventing the rise of the total temperature level of the LHP \citep{Ku2008}. 

The operating range of the considered LHP in this paper is defined as follows: 
\begin{align}
20\,\text{W}\leq\dot{Q}_{evap}&\leq 100\,\text{W},\\	
-15\,^\circ\text{C}\leq\hspace{\halfdiffT}T_{sink}\hspace{\halfdiffT}&\leq 15\,^\circ\text{C},\\	
0\,\text{W}\leq\hspace{\halfdiffQ}\dot{Q}_{cc}\hspace{\halfdiffQ}&\leq 10\,\text{W},\label{eq:heater}\\
T_{set}\hspace{\halfdiffTset}&=27\,^\circ\text{C},\label{eq:Tset}\\
T_{amb}\hspace{\halfdiffTamb}&=25\,^\circ\text{C}.\label{eq:amb}
\end{align}
The operating range includes only the \textit{variable conductance mode}, since the condenser is never fully used. The considered LHP is equipped with an electrical control heater on the CC providing a heat input $\dot{Q}_{cc}$ to the CC. The control heater enables the suppression of the disturbances $\dot{Q}_{evap}$ and $T_{sink}$, which influence the curvature of the natural SSOT in Fig.~\ref{Abb:op_charac}. Thus, this paper focuses on the LHP behavior for heat loads between $\dot{Q}_{low}$ and $\dot{Q}_{trans}$.

\section{LHP state-space modeling}
\label{sec:res}
In this section, an analytical transient model of the LHP with lumped parameters is created in the form of a state-space model as required for model-based control design. In contrast to the previous work in \cite{Gellrich2018b}, the aim in this paper is to model the fluid's dynamics to extend the scope and to improve the accuracy of the LHP model. Furthermore, a detailed modeling of the condenser and the LL is included. In order to ease notation, all dependencies of the variables are dropped in the following equations. 

\begin{figure}[htb]
	\centering
	\resizebox{1.00\linewidth}{!}{\input{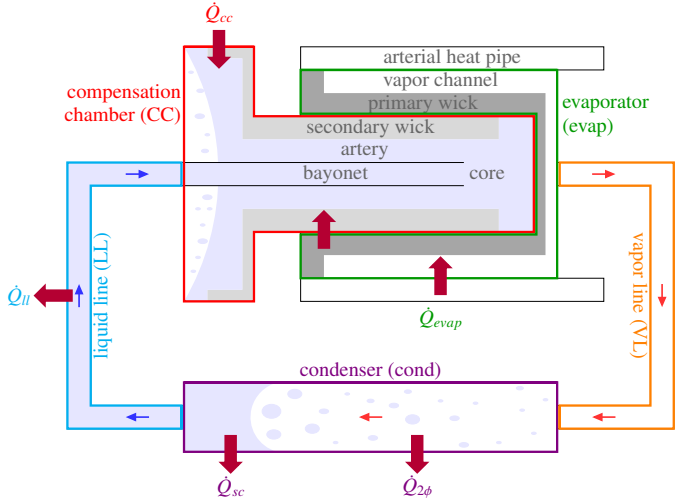}}
	\caption{Representation of the five LHP subsystems}	
	\label{Abb:Subsysteme}
\end{figure}

To derive the basic thermodynamic equations of the LHP, the system is divided into subsystems used as control volumes for the conservation laws. The five subsystems, presented in Fig.~\ref{Abb:Subsysteme}, are the compensation chamber (CC) in red, the evaporator (evap) in green, the vapor line (VL) in orange, the liquid line (LL) in blue, and the condenser (cond) in violet.

Before the individual equations of the subsystems are established, general assumptions for modeling the LHP are stated, which are necessary to simplify the system without changing the basic system's dynamics.
\begin{assumption}
	\label{ass:ideal}
	The vapor is assumed to be an ideal gas, the liquid is incompressible.
\end{assumption}
\begin{assumption}
	\label{ass:super}
	No superheating is regarded, since little superheating takes place in the considered operating range of the LHP (see \cite{Meinicke2019}). In addition, the length of the superheated region in the condenser is negligibly small compared to the lengths of the two-phase region and the subcooled region (see also \cite{Launay2007}).
\end{assumption}
\begin{assumption}
	\label{ass:sat}
	In the CC, the evaporator, and the two-phase region of the condenser, the fluid is in saturated state at all times.
\end{assumption}
\begin{assumption}
	\label{ass:sink}
	The sink temperature is considered to be spatially constant, since the temperature gradient along the condenser length varies only by a few degrees Kelvin as experimental data from the LHP test bench of \cite{Meinicke2019} show.
\end{assumption}
\begin{assumption}
	\label{ass:inertia}
	For the single-phase subsystems, which are the evaporator and the transport lines, the thermal and the fluid inertia are neglected assuming that neither heat nor mass are stored in the subsystem, since the corresponding temperature measurements show almost no delay between the inlet and the outlet of the subsystems.
\end{assumption}
\begin{assumption}
	\label{ass:physic}
	The physical properties of the working fluid are functions of the respective subsystem temperatures in the operating point. Their temperature derivatives, except for the vapor density's derivative, are neglected, as their temperature dependency is small.
\end{assumption}
\begin{assumption}
	\label{ass:flow}
	In the considered operating range of the LHP, the flow regime of the working fluid in the loop is laminar.
\end{assumption}

\subsection{Compensation chamber}

The CC subsystem includes the secondary wick and the evaporator core in order to set the boundary between the CC and the evaporator at the liquid-vapor interface inside the primary wick as depicted in Fig.~\ref{Abb:Subsysteme}. During LHP operation, the primary and the secondary wick are assumed to be soaked with liquid at all times, i.e. the fluid enters and leaves the CC in liquid state. The CC contains both a liquid and a vapor phase in saturated states. In order to model the CC, the approach of \cite{Launay2007} is applied. In this approach, the mass and the energy balance are evaluated for the fluid inside the CC.

The mass balance implies that the change of mass contained in the CC equals the difference of the inlet and the outlet mass flow. Therefore, the mass balance is described by the following equation:
\begin{equation}
	\frac{d}{dt}\left\{m_{cc}^l+m_{cc}^v\right\}=\dot{m}_l-\dot{m}_v,
	\label{eq:Massenerh.CC1}
\end{equation}
where $\dot{m}_l$ denotes the liquid mass flow, which enters the CC and $\dot{m}_v$ denotes the mass flow that evaporates at the outer surface of the primary wick. The liquid and the vapor mass in the CC are expressed as
\begin{align}
m_{cc}^{l}&=\rho_{cc}^{l}\cdot V_{cc}^{l},\label{eq:mccl}\\
m_{cc}^{v}&=\rho_{cc}^{v}\cdot V_{cc}^{v}\label{eq:mccv}
\end{align}
with the corresponding densities $\rho_{cc}^{l}$ and $\rho_{cc}^{v}$, and volumes $V_{cc}^l$ and $V_{cc}^v$, for which hold:
\begin{equation}
V_{cc}=V_{cc}^l+V_{cc}^v.
\label{eq:Vcc}
\end{equation}
By applying the ideal gas law (see Assumption~\ref{ass:ideal})
\begin{equation}
\label{eq:idealGasLaw}
\rho_v = \frac{p}{R T}
\end{equation} 
to the vapor density and the Clausius-Clapeyron-equation
\begin{equation}
\label{eq:ClausiusClapeyron}
\frac{dp}{dT} = \frac{\Delta h_v}{T \left(\frac{1}{\rho_v}-\frac{1}{\rho_l}\right)}
\end{equation}
to the temperature derivative $\frac{dp}{dT}$ of the pressure $p$, the time derivative of the vapor density is: 
\begin{equation}
\label{eq:rhovdt}
\frac{d\rho_v}{dt} = \left(\frac{\Delta h_v \: \rho_l \: \rho_v}{R \: T^2 \: (\rho_l-\rho_v)}-\frac{\rho_v}{T}\right)\cdot\frac{dT}{dt}
\end{equation}
with the specific heat of evaporation $\Delta h_v$ and the specific gas constant $R$. Equations \eqref{eq:mccl}, \eqref{eq:mccv}, \eqref{eq:Vcc}, and \eqref{eq:rhovdt} lead to the following form of \eqref{eq:Massenerh.CC1}:
\begin{align}
\label{eq:Massenerh.CC}
&\hphantom{=\:}\left(\rho_{cc}^l-\rho_{cc}^v\right)\cdot\frac{dV_{cc}^{l}}{dt}\nonumber\\
&+\left(V_{cc}-V_{cc}^{l}\right)\cdot\left(\frac{\Delta h_{cc}^{v}\rho_{cc}^l\rho_{cc}^v}{RT_{cc}^{2}(\rho_{cc}^l-\rho_{cc}^v)}-\frac{\rho_{cc}^v}{T_{cc}}\right)\cdot\frac{dT_{cc}}{dt}\nonumber\\
&=\dot{m}_{l}-\dot{m}_{v},
\end{align}
which is in accordance with \cite{Launay2007}. The change of mass in the CC in \eqref{eq:Massenerh.CC} is a function of the volume $V_{cc}^l$ of the liquid inside the CC, the homogeneous saturation temperature $T_{cc}$ in the CC, and their change rates respectively.

The thermal energy balance of the CC is given by
\begin{align}
\label{eq:Energieerh.CC1}
&\hphantom{=\:\:}\frac{d}{dt}\left\{\rho_{cc}^l h_{cc}^l V_{cc}^l\right\}+\frac{d}{dt}\left\{\rho_{cc}^v h_{cc}^v V_{cc}^v\right\}-\frac{d}{dt}\left\{p_{cc} V_{cc} \right\}\nonumber\\
&=\dot{m}_l h_{cc,in}^l-\dot{m}_v h_{cc}^l+\dot{Q}_{cc}+\dot{Q}_{wick}.
\end{align}
In \eqref{eq:Energieerh.CC1}, the left-hand side describes the change of thermal energy in the CC, which consists of the enthalpy of the liquid and the vapor phase minus the flow work required for pushing the fluid out of the CC. The right-hand side of \eqref{eq:Energieerh.CC1} describes the energy flows across the boundary of the subsystem. The enthalpy flows in and out of the CC depend on the corresponding mass flow rates $\dot{m}_{l}$ and $\dot{m}_{v}$. The heat flow $\dot Q_{cc}$ is applied to the CC by the control heater, which is installed on the CC. $\dot Q_{wick}$ denotes the heat flow, which leaks from the evaporator to the CC by heat conduction through the wicks. It is induced by the temperature difference between the two subsystems. According to \cite{Ku1999}, this leaking heat flow is given by 
\begin{equation}
\label{eq:Qwick}
\dot Q_{wick}=\frac{1}{R_{wick}} \cdot \left(T_{evap}-T_{cc}\right)
\end{equation} 
with the thermal resistance $R_{wick}$ of the wicks.

By inserting \eqref{eq:Vcc}, the ideal gas law \eqref{eq:idealGasLaw}, the Clausius-Clapeyron equation \eqref{eq:ClausiusClapeyron}, and the mass balance \eqref{eq:Massenerh.CC} into \eqref{eq:Energieerh.CC1}, the energy balance of the CC can be rewritten as:
\begin{align}
\label{eq:Energieerh.CC}
&\hphantom{=\:}\left[\left(\rho_{cc}^l V_{cc}^l+\left(V_{cc}-V_{cc}^l\right)\cdot\rho_{cc}^v\right)\cdot c_{p,cc}^l\vphantom{\frac{V_{cc}^l}{V_{cc}^l}}\right.\nonumber\\
&+\Delta h_{cc}^v\cdot\left(V_{cc}-V_{cc}^l\right)\cdot\left(\frac{\Delta h_{cc}^v \rho_{cc}^l\rho_{cc}^v }{R T_{cc}^2(\rho_{cc}^l-\rho_{cc}^v)}-\frac{\rho_{cc}^v }{T_{cc}}\right)\nonumber\\
&-V_{cc}\cdot\frac{\Delta h_{cc}^v\rho_{cc}^l\rho_{cc}^v}{T_{cc}(\rho_{cc}^l-\rho_{cc}^v)}\left.\vphantom{\frac{V_{cc}^l}{V_{cc}^l}}\right]\cdot\frac{dT_{cc}}{dt}-\rho_{cc}^v\Delta h_{cc}^v\cdot\frac{dV_{cc}^l}{dt}\nonumber\\
&=\dot{m}_l c_p^l\cdot(T_{cc,in}-T_{cc})+\dot{Q}_{cc}+\dot{Q}_{wick}.
\end{align}
The specific heat capacity $c_p^l$ of the liquid phase on the right-hand side of \eqref{eq:Energieerh.CC} is obtained by the arithmetic mean of $c_p^l(T_{cc,in})$ and $c_p^l(T_{cc})$ neglecting the small temperature dependency of the specific heat capacity (see Assumption~\ref{ass:physic}). $T_{cc,in}$ denotes the temperature at the inlet of the CC.

The simultaneous differential equations \eqref{eq:Massenerh.CC} and \eqref{eq:Energieerh.CC} are rearranged resulting in the time derivatives of $T_{cc}$ and $V_{cc}^l$:
\begin{align}
\frac{dT_{cc}}{dt}&=\frac{\dot{m}_l c_p^l\left(T_{cc,in}-T_{cc}\right)+\dot{Q}_{cc}+\dot{Q}_{wick}+A\left(\dot{m}_l-\dot{m}_v\right)}{\left(\rho_{cc}^l  V_{cc}^l+D\rho_{cc}^v\right)c_{p,cc}^l +CD\,\Delta h_{cc}^v-\frac{V_{cc}R}{T_{cc}}B+ACD},\label{eq:Tcc}\\
\frac{dV_{cc}^l}{dt}&=\frac{1}{\rho_{cc}^{l}-\rho_{cc}^{v}}\cdot\left[\dot{m}_{l}-\dot{m}_{v}-CD\vphantom{\frac{V_{cc}^l}{V_{cc}^l}}\right.\nonumber\\
&\;\cdot\frac{\dot{m}_l c_p^l\left(T_{cc,in}-T_{cc}\right)+\dot{Q}_{cc}+\dot{Q}_{wick}+A\left(\dot{m}_l-\dot{m}_v\right)}{\left(\rho_{cc}^l  V_{cc}^l+D\rho_{cc}^v\right)c_{p,cc}^l+CD\,\Delta h_{cc}^v-\frac{V_{cc}R}{T_{cc}}B+ACD} \label{eq:Vccl}\left.\vphantom{\frac{V_{cc}^l}{V_{cc}^l}}\right]
\end{align}
with 
\begin{align}
\begin{split}
A&= \frac{\rho_{cc}^v\Delta h_{cc}^v}{\rho_{cc}^l-\rho_{cc}^v},\\
B&=\frac{\rho_{cc}^l}{R}\cdot A,\\
C&=\frac{1}{T_{cc}^2}\cdot B-\frac{\rho_{cc}^v}{T_{cc}},\\
D&=V_{cc}-V_{cc}^l.\label{eq:ABCD}
\end{split}
\end{align}

\subsection{Evaporator}
\label{sec:Evaporator}

As stated in Sec.~\ref{sec:fund}, the liquid leaving the CC evaporates completely at the outer surface of the primary wick so that only vapor is present inside the evaporator. According to Assumptions~\ref{ass:super} and \ref{ass:sat}, superheating is neglected and the vapor is assumed to be saturated at all times. In this paper, the evaporator saturation temperature is modeled as a function of the CC temperature and the pressure difference between the CC and the evaporator, because the energy balance of the evaporator control volume is used to calculate the vapor mass flow rate as described in Sec. \ref{sec:Transportlines}.

In saturated state, temperature and pressure are linked to each other. The dependency is described by the Antoine equation \citep{Antoine1888} for the pressure in the CC as
\begin{equation}
\label{eq:Antoine_Druck}
p_{cc}=10^{A_{wf}-\frac{B_{wf}}{C_{wf}+T_{cc}}},
\end{equation}
and the temperature $T_{evap}$, respectively, as
\begin{equation}
\label{eq:Antoine_Temp}
T_{evap}=-\frac{B_{wf}}{\log_{10}(p_{evap})-A_{wf}}-C_{wf}, 
\end{equation}
where $A_{wf}$, $B_{wf}$, and $C_{wf}$ are specific parameters of the working fluid (see \ref{sec:appendix}). The pressure difference between the CC and the evaporator is given by the pressure difference $\Delta p_{cap}$ caused by capillary forces in the pores of the primary wick minus the pressure losses $\Delta p_{fri}$ inside the wick:
\begin{equation}
\label{eq:pevap}
p_{evap}=p_{cc}+\Delta p_{cap}-\Delta p_{fri}.
\end{equation}
The formula for maximum capillary pressure is established by putting the minimum curvature radius, i.e. the wick pore radius $R_p$, and a minimum contact angle of $\theta=\SI{0}{\degree}$ into the Young-Laplace equation for tubes (see \cite{Ku1999}): 
\begin{equation}
\Delta p_{cap}=\frac{2\sigma \cos\left(\theta\right)}{R_p}
\label{eq:MaxKapillardruck}
\end{equation}
with the surface tension $\sigma$ of the saturated liquid working fluid (see \ref{sec:appendix}). As the pore radius $R_p$ of the wick in the considered LHP is unknown, it is chosen as $R_p= \SI{1}{\micro \meter}$ by means of literature values (\citep{Chuang2003}, \citep{Kaya2006}, \citep{Ren2007}).

Following Assumption~\ref{ass:inertia}, the equation to describe the vapor mass flow rate is derived from the steady-state energy balance of the evaporator and given by
\begin{equation}
\label{eq:mv}
\dot{m}_{v}=\frac{\dot{Q}_{evap}-\dot{Q}_{wick}}{c_p^l\left(T_{evap}-T_{cc}\right)+\Delta h_{evap}^v}
\end{equation}
with the specific heat capacity $c_p^l$ of the liquid phase as the arithmetic mean of $c_p^l(T_{evap})$ and $c_p^l(T_{cc})$. The heat load $\dot{Q}_{evap}$ is applied to the evaporator via the arterial heat pipes. A part of the heat load forms the heat leak $\dot{Q}_{wick}$ between the evaporator and the CC (see \eqref{eq:Qwick}). The residual heat $\left(\dot{Q}_{evap}-\dot{Q}_{wick}\right)$ in \eqref{eq:mv} causes the evaporation of the liquid at the outer surface of the wick.

\subsection{Transport lines}
\label{sec:Transportlines}

The presented LHP model describes the LHP behavior for sink temperatures lower than the ambient temperature (see Sec.~\ref{sec:fund}). The evaporator temperature near the heat load is the highest temperature in the LHP and follows the CC temperature, which is controlled near the ambient temperature (see \eqref{eq:Tset} and \eqref{eq:amb}). Thus, the temperature difference between the VL and the surroundings is small and for this reason a heat exchange between the VL and the surroundings is neglected. According to Assumption~\ref{ass:super}, no superheating arrives at the condenser inlet and the vapor enters the condenser in saturated state.

While the heat exchange with the surroundings at the VL is neglected, the heat exchange at the LL must be considered. The condenser outlet temperature $T_{cond,out}$ is the lowest temperature of the LHP and close to the sink temperature when the LHP operates in variable conductance mode (see Sec.~\ref{sec:fund}). The temperature difference between the ambient temperature and the LL temperature determines the behavior of the operating temperature, i.e. the characteristic U-shaped curve of the SSOT. In accordance with \cite{Meinicke2019}, the LL is modeled as a heat exchanger with a fixed ambient temperature $T_{amb}$ (see \eqref{eq:amb}). The thermal and the fluid inertia of the liquid in the LL are neglected (see Assumption~\ref{ass:inertia}) and the temperature decreases exponentially along the LL. The LL outlet temperature, which corresponds to the CC inlet temperature $T_{cc,in}$, is given by the following equation:
\begin{equation}
\label{eq:Tccin}
T_{cc,in}=T_{amb}+\left(T_{cond,out}-T_{amb}\right)\cdot\exp\left(-\frac{\pi D_{c}k_{ll}L_{ll}}{\dot{m}_{l}c_{p}^{l}}\right),
\end{equation}
where $k_{ll}$ denotes the heat transfer coefficient of the LL, $L_{ll}$ the length of the LL, and $D_{c}$ the diameter of the LL. The mean specific heat capacity $c_p^l$ in \eqref{eq:Tccin} is obtained by the arithmetic mean of $c_p^l(T_{cc,in})$ and $c_p^l(T_{cond,out})$.

\subsection{Condenser}
\label{sec:Condenser}

As stated in Sec.~\ref{sec:Transportlines}, the vapor flow enters the condenser from the VL in saturated state. For this reason, the condenser is divided into a saturated, two-phase region and a subcooled, single-phase region. It is assumed that the fluid enters the two-phase region as pure vapor. Furthermore, the temperature $T_{cond}$ and the pressure $p_{cond}$ in the two-phase region are assumed to be homogeneous, as pure substances condense at constant pressure and temperature. In order to model the dynamic behavior of the length of the two-phase region, the condensation model of \cite{Wedekind1978}, which is also used in \cite{Launay2007}, is applied to the condenser of the LHP. The zero-dimensional, transient condensation model provides an interface tracking by describing the point of complete condensation and the liquid mass flow rate at the outlet of the condenser as a function of the vaporous mass flow rate at the inlet of the condenser. The advantage of this model is that it is a dynamic, lumped-parameter model that needs only one constant, which is the heat transfer coefficient of the two-phase region $k_{2\phi}$. The behavior of the condenser is assumed to be governed by mass and energy conservation principles, while the influence of momentum and thermal inertia are neglected. The distance between the condenser inlet and the point of complete condensation is described by $\eta$, which denotes the length of the two-phase region in the condenser as part of the total condenser length $L_{cond}$, as shown in Fig. \ref{Abb:Kondensator}.
\begin{figure}[htb]
	\centering
	\resizebox{1.00\linewidth}{!}{\usetikzlibrary{patterns}
\usetikzlibrary{shapes.arrows}
\begin{tikzpicture}
	\tikzstyle{every node}=[font={\fontsize{20}{0}\selectfont}]
	\tikzstyle{<->} = [thick, triangle 60-triangle 60];
	
	\definecolor{redish}{RGB}{255,50,50}
	\definecolor{bluish}{RGB}{50,50,255}
	\definecolor{red}{RGB}{255,0,0}
	\definecolor{green}{RGB}{0,153,0}
	\definecolor{violet}{RGB}{128,0,128}
	\definecolor{greenish}{RGB}{0,176,80}
	\definecolor{cyan}{RGB}{0,176,240}
	\definecolor{orange}{RGB}{255,128,0}	
	\definecolor{darkRed}{RGB}{180,0,50}
	
	\draw[blue!10,fill] (7.75,2.25) ellipse (3mm and 2mm);
	\draw[blue!10,fill] (8.75,0.5) ellipse (3mm and 2mm);
	\draw[blue!10,fill] (9,2) ellipse (2mm and 1mm);
	\draw[blue!10,fill] (7.5,1.25) ellipse (2mm and 1mm);
	\draw[blue!10,fill] (10,1) ellipse (2mm and 1mm);
	\draw[blue!10,fill] (9,1.25) ellipse (1.5mm and 0.75mm);
	\draw[blue!10,fill] (10.1,2.4) ellipse (1.5mm and 0.75mm);
	\draw[blue!10,fill] (12,2.7) ellipse (1.5mm and 0.75mm);
	\draw[blue!10,fill] (14,2.2) ellipse (1mm and 0.5mm);
	\draw[blue!10,fill] (13.4,0.25) ellipse (1.5mm and 0.75mm);
	\draw[blue!10,fill] (11.4,1.6) ellipse (1.5mm and 0.75mm);
	\draw[blue!10,fill] (15.5,0.75) ellipse (1mm and 0.5mm);
	\draw[blue!10,fill] (12.5,0.9) ellipse (1mm and 0.5mm);
	
	\fill[blue!10] (0,3) -- (0,0) -- (8,0) arc (75:-75:-1.55) ;
	\draw[line width=2.8pt,violet] (0,0) -- (0,3) -- (16,3) -- (16,0) -- (0,0) -- (0,3);
	\draw[<->] (0,5.5) -- node[midway,above,xshift=0mm] {$L_{cond}$} (16,5.5);
	\draw[<->] (6.85,4.5) -- node[midway,above,xshift=0mm] {$\eta$} (16,4.5);
	
	\node[] at (3.425,3.5) {subcooled region};
	\node[] at (11.425,3.5) {two-phase region};
	
	\draw[-triangle 45, line width=2pt, redish] (18,1.5) -- node[midway,above,xshift=0mm,black] {$\dot{m}_{v}$} (16,1.5);
	\draw[-triangle 45, line width=2pt, bluish] (0,1.5) -- node[midway,above,xshift=0mm,black] {$\dot{m}_{l}$} (-2,1.5);
	\node[right color=redish,left color=bluish,shading angle=90,single arrow,rotate=180,minimum height=2.01cm,line width=0.01mm,
	single arrow head extend=0.16cm,single arrow tip angle=45,single arrow head indent=0cm,inner sep=1pt] at (5.88,1.5) {};
	\node[yshift=5mm,xshift=0mm] at (5.88,1.5) {$\dot{m}^*$};
	
	\node[draw=darkRed,fill=darkRed,single arrow,rotate=270,minimum height=1.5cm,single arrow head extend=0.15cm,line width=1.5mm] at (3.425,0) {};
	\node[violet] at (3.425,-1.5) {$\dot{Q}_{sc}$};
	\node[draw=darkRed,fill=darkRed,single arrow,rotate=270,minimum height=1.5cm,single arrow head extend=0.15cm,line width=1.5mm] at (11.425,0) {};
	\node[violet] at (11.425,-1.5) {$\dot{Q}_{2\phi}$};
	\node[circle,fill=bluish,inner sep=3pt] at (0,1.5) {};
	\node[black,below] at (1.5,1.9) {$T_{cond,out}$};
	\node[circle,fill=redish,inner sep=3pt] at (6.85,1.5) {};
	\node[black,below] at (7.91,1.9) {$T_{cond}$};
	\node[circle,fill=redish,inner sep=3pt] at (16,1.5) {};
	\node[black,below] at (14.95,1.9) {$T_{cond}$};
		
\end{tikzpicture}}
	\caption{Schematic of the condenser}	
	\label{Abb:Kondensator}
\end{figure}
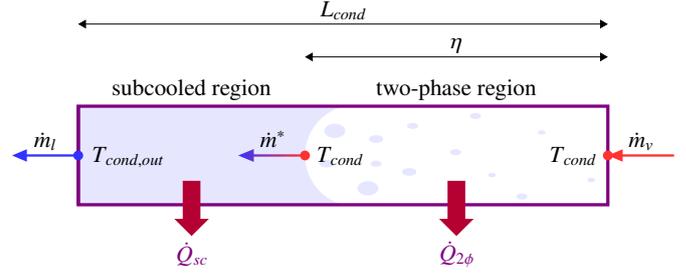\newline
The mass balance for the two-phase region yields
\begin{equation}
	\label{eq:Massenerh.2phi}
	\frac{d}{dt}\left\{\rho_{2\phi}\cdot A_c\cdot\eta\right\}=\dot{m}_v-\dot{m}^*
\end{equation}
with the cross-sectional area $A_c$ of the condenser, the inlet mass flow rate $\dot{m}_v$ and the outlet mass flow rate $\dot{m}^*$. The variable $\dot{m}^*$ denotes the mass flow rate, which flows across the moving interface between the two-phase region and the subcooled region (see Fig. \ref{Abb:Kondensator}). In order to simplify the problem, the condenser model assumes a homogeneous density in the two-phase region:
\begin{equation}
\label{eq:rho_2phi}
\rho_{2\phi}= \rho_{cond}^l\cdot\left(1-\bar{\alpha}\right)+\rho_{cond}^v\cdot\bar{\alpha}.
\end{equation}
It is determined by the so-called mean void fraction $\bar{\alpha}$, which is time invariant and hence neglects the transient changes of the fluid momentum. According to \cite{Wedekind1978}, a value of
\begin{equation}
 \bar{\alpha}=0.82
\end{equation}
is chosen for the condensation in the LHP.

According to the first law of thermodynamics, the energy balance of the two-phase region is
\begin{align}
\label{eq:Energieerh.2phi}
&\hphantom{=\:\:}\frac{d}{dt}\left\{\left[\rho_{cond}^l\cdot h_{cond}^l\cdot\left(1-\bar{\alpha}\right)+\rho_{cond}^v\cdot h_{cond}^v\cdot\bar{\alpha}\right]\cdot A_c\cdot\eta\right\}\nonumber\\
&=-\dot{Q}_{2\phi}+h_{cond}^v\cdot\dot{m}_v-h_{cond}^l\cdot\dot{m}^*,
\end{align}
where the left-hand side describes the change of homogeneous enthalpy in the subsystem and the right-hand side stands for the enthalpy flows of the entering vapor and the leaving condensate minus the heat flow $\dot{Q}_{2\phi}$. The variable $\dot{Q}_{2\phi}$ denotes the heat flow, which is released during condensation and is transferred to the heat sink by heat convection and heat conduction. It is expressed by
\begin{equation}
\dot{Q}_{2\phi}=k_{2\phi}\cdot\pi\cdot D_c\cdot\eta\cdot\left(T_{cond}-T_{sink}\right)
\end{equation}
with the aforementioned heat transfer coefficient of the two-phase region $k_{2\phi}$ and the diameter $D_{c}$ of the condenser. The sink temperature $T_{sink}$ is assumed to be spatially constant along the condenser (see Assumption~\ref{ass:sink}).

By combining \eqref{eq:Massenerh.2phi} and \eqref{eq:Energieerh.2phi}, the differential equation for $\eta$ is obtained \citep{Wedekind1978}:
\begin{equation}
	\label{eq:eta}
	\frac{d\eta}{dt}=-\frac{k_{2\phi}\pi D_{c}(T_{cond}-T_{sink})}{\rho_{cond}^{v}\bar{\alpha}A_{c}\Delta h_{cond}^v}\cdot\eta+\frac{1}{\rho_{cond}^v\bar{\alpha}A_{c}}\cdot\dot{m}_{v}.
\end{equation}
The equation for the mass balance of the subcooled region 
\begin{equation}
	\label{eq:Massenerh.sc}
	\frac{d}{dt}\left\{\rho_{cond}^l\cdot A_c\cdot\left(L_{cond}-\eta\right)\right\}=\dot{m}^*-\dot{m}_l,
\end{equation}
together with \eqref{eq:Massenerh.2phi} and \eqref{eq:Energieerh.2phi} is used to express the mass flow rate $\dot{m}^*$ \citep{Launay2007}:
\begin{equation}
\label{eq:m_Stern}
\dot{m}^*=\dot{m}_v-\frac{\dot{m}_v-\dot{m}_l}{1-\frac{\rho_{cond}^l}{\rho_{2\phi}}}.
\end{equation}
In order to determine the saturation temperature $T_{cond}$, the steady-state energy balance of the two-phase region is evaluated:
\begin{equation}
	\label{eq:T_cond}
	T_{cond}=T_{sink}+\frac{\Delta h_{cond}^v}{k_{2\phi}\pi D_{c}\eta}\cdot\dot{m}^*.
\end{equation} 
While the temperature $T_{cond}$ of the two-phase region is assumed to be spatially constant, the temperature of the liquid in the subcooled region decreases along the length $\left(L_{cond}-\eta\right)$ due to the heat flow $\dot{Q}_{sc}$ to the heat sink. The subcooling process is modeled by describing the subcooled region as a heat exchanger at a constant casing temperature $T_{sink}$, analogously to the description of $T_{cc,in}$ in the LL. This results in the condenser outlet temperature
\begin{align}
\label{eq:T_condout}
T_{cond,out}&=T_{sink}+\left(T_{cond}-T_{sink}\right)\nonumber\\
&\;\cdot\exp\left[-\frac{\pi D_{c}k_{sc}}{\dot{m}_{l}c_{p}^{l}}\cdot\left(L_{cond}-\eta\right)\right]
\end{align}
with the heat transfer coefficient $k_{sc}$ of the subcooled region. The heat capacity $c_p^l$ in \eqref{eq:T_condout} is described by the arithmetic mean of $c_p^l(T_{cond})$ and $c_p^l(T_{cond,out})$.

\subsection{Liquid mass flow rate}

For the description of the liquid mass flow rate, the transient form of the conservation of momentum of the liquid is evaluated. The considered control volume consists of the liquid fluid column in the liquid line and the condenser. Therewith, the differential equation for the incompressible liquid mass flow rate $\dot{m}_l$ is obtained (see Assumption~\ref{ass:ideal}): 
\begin{align}
	\label{eq:m_l}
	\frac{d\dot{m}_{l}}{dt}&=\frac{1}{L_{cond}-\eta+L_{ll}}\cdot\left(\frac{\dot{m}_{l}^2}{\rho_{cond}^l A_{c}}-\frac{\dot{m}_{l}^2}{\rho_{cc,in}^l A_{c}}+p_{cond}\cdot A_{c}\right.\nonumber\\
	&-p_{cc}\cdot A_{c}-\frac{128}{\pi}\cdot\frac{L_{cond}-\eta+L_{ll}}{D_{c}^{4}}\cdot\frac{\mu_{l}A_{c}\dot{m}_{l}}{\rho_{cc,in}^{l}}\left.\vphantom{\frac{\dot{m}_{l}^2}{\rho_{cc,in}^l A_{c}}}\right).
\end{align}
The pressures $p_{cc}$ and $p_{cond}$ are deduced from the respective saturation temperatures by using the Antoine equation~\eqref{eq:Antoine_Druck}. The occurring pressure losses are calculated in accordance with Assumption~\ref{ass:flow}. Eq. \eqref{eq:m_l} is similar to the equation used in \cite{Launay2007} without LHP inclination. To simplify the equation, the total liquid column between the vapor-liquid interfaces in the condenser and the CC is calculated without the liquid length in the CC, since the liquid length in the CC is relatively small in comparison with the liquid lengths in the LL and the condenser. In addition, the dynamic influence of the vapor-liquid interface in the CC on the liquid column is negligible compared to the other terms in \eqref{eq:m_l}.

It should be noted that in contrast to the depiction in Fig.~\ref{Abb:Subsysteme}, the condenser and the transport lines have the same diameter $D_c$ and therefore the same cross-sectional area $A_c$.

\subsection{LHP state-space model}

After completing the thermodynamic modeling of the LHP, the state-space model is derived. By inserting the equations \eqref{eq:Qwick}, \eqref{eq:Antoine_Druck}, \eqref{eq:Antoine_Temp}, \eqref{eq:pevap}, \eqref{eq:mv}, \eqref{eq:Tccin}, \eqref{eq:T_cond}, and \eqref{eq:T_condout} into the four differential equations \eqref{eq:Tcc}, \eqref{eq:Vccl}, \eqref{eq:eta}, and \eqref{eq:m_l}, the dynamical behavior of the LHP is described by a nonlinear parameter-variant state-space model:
\begin{align}
 \label{eq:ss}
 \begin{split}
 \dot{\boldsymbol{x}}&=
 \frac{d}{dt}
 \begin{bmatrix}
  T_{cc}\\
  \eta\\
  V_{cc}^l\\
  \dot{m}_l
 \end{bmatrix}
 =
 \boldsymbol{f}\left(T_{cc},\eta,V_{cc}^l,\dot{m}_l,\dot{Q}_{evap},T_{sink},\dot{Q}_{cc}\right),\\
 y&=T_{cc}
 \end{split}
\end{align}
with the variable lumped parameters $R_{wick}$, $k_{2\phi}$, $k_{sc}$, and $k_{ll}$.

\subsection{Parameter determination}
\label{sec:param}

The LHP as thermodynamic system is a nonlinear parameter-variant system, where the parameters further depend on flow regimes, velocity-dependent pressure losses or spatial variation. Hence, the lumped parameters of \eqref{eq:ss} are determined in a chosen operating point (OP) for the subsequent model-based control design in the following section.

The temperature-dependent parameters are calculated from the corresponding temperatures in the OP with the functions in \ref{sec:appendix} (see Assumption~\ref{ass:physic}). Following the parameter determination process in \cite{Gellrich2018b}, the lumped parameters $R_{wick}$, $k_{2\phi}$, $k_{sc}$, and $k_{ll}$ are calculated by solving the stationary equations of \eqref{eq:ss}. Unfortunately, the system of equations for the lumped parameters is not solvable due to the dependent differential equations \eqref{eq:Tcc} and \eqref{eq:Vccl}, and the implicit forms of \eqref{eq:Tccin}, \eqref{eq:T_cond}, and \eqref{eq:T_condout}. For the solution of this problem, the temperatures $T_{cond,out}^{op}$ and $T_{cond}^{op}$ must be predefined with their values in the OP, which can be derived from experimental data. Hence, the new system of equations contains the differential equations \eqref{eq:Tcc} and \eqref{eq:eta}, and the equations \eqref{eq:T_cond}, \eqref{eq:T_condout}, and \eqref{eq:Tccin}. The solution of this system includes the four lumped parameters ($R_{wick}^{op}$, $k_{2\phi}^{op}$, $k_{sc}^{op}$, and $k_{ll}^{op}$) and the CC inlet temperature ($T_{cc,in}^{op}$) in the OP.

\section{Model-based control design}
\label{sec:control}
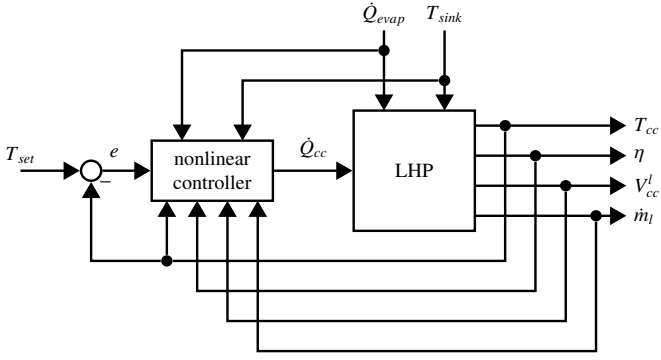
\begin{figure}[htb]
	\centering
	\resizebox{1.00\linewidth}{!}{\begin{tikzpicture}
\tikzstyle{->} = [very thick, -triangle 60];

\tikzstyle{block} = [draw,minimum height=30, minimum width=60,very thick];
\tikzstyle{add} = [draw,circle,minimum size=10,very thick];
\tikzstyle{quadrat} = [draw,minimum height=60, minimum width=60,very thick];
\tikzstyle{point} = [fill,circle,inner sep=2];

\definecolor{gold}{RGB}{255,192,0}
\definecolor{green}{RGB}{0,176,80}
\definecolor{blue}{RGB}{0,112,192}


\node[add] (add1) {};
\coordinate[left of=add1,node distance=35] (coo1);
\coordinate[below of=add1,node distance=45] (coo2);
\node[block,right of=add1,node distance=60,color=black,align=center] (Ly) {nonlinear\\controller};
\coordinate[right of=Ly,node distance=45] (coo4);
\node[quadrat,right of=coo4,node distance=55] (lhp) {LHP};
\draw[->] (Ly) -- node[midway,above,yshift=1mm] {$\dot{Q}_{cc}$} (lhp);
\coordinate[below of=coo4,node distance=90] (coo4b);
\coordinate[right of=lhp,node distance=30] (coo5m);
\coordinate[above of=coo5m,node distance=22.5] (coo5a);
\coordinate[above of=coo5m,node distance=7.5] (coo5b);
\coordinate[below of=coo5m,node distance=7.5] (coo5c);
\coordinate[below of=coo5m,node distance=22.5] (coo5d);
\coordinate[right of=coo5a,node distance=75] (coo5aa);
\coordinate[right of=coo5b,node distance=75] (coo5bb);
\coordinate[right of=coo5c,node distance=75] (coo5cc);
\coordinate[right of=coo5d,node distance=75] (coo5dd);
\node[point,right of=coo5a,node distance=15] (coo5aap) {};
\node[point,right of=coo5b,node distance=30] (coo5bbp) {};
\node[point,right of=coo5c,node distance=45] (coo5ccp) {};
\node[point,right of=coo5d,node distance=60] (coo5ddp) {};
\coordinate[below of=coo5aap,node distance=67.5] (coo5aapp);
\coordinate[below of=coo5bbp,node distance=67.5] (coo5bbpp);
\coordinate[below of=coo5ccp,node distance=67.5] (coo5ccpp);
\coordinate[below of=coo5ddp,node distance=67.5] (coo5ddpp);
\draw[->] (coo5a) -- node[at end,right,yshift=0mm] {$T_{cc}$} (coo5aa);
\draw[->] (coo5b) -- node[at end,right,yshift=0mm] {$\eta$} (coo5bb);
\draw[->] (coo5c) -- node[at end,right,yshift=0mm] {$V_{cc}^l$} (coo5cc);
\draw[->] (coo5d) -- node[at end,right,yshift=0mm] {$\dot{m}_l$} (coo5dd);
\coordinate[below of=Ly,node distance=15] (Lym);
\coordinate[left of=Lym,node distance=22.5] (coo7a);
\coordinate[left of=Lym,node distance=7.5] (coo7b);
\coordinate[right of=Lym,node distance=7.5] (coo7c);
\coordinate[right of=Lym,node distance=22.5] (coo7d);
\node[point,below of=coo7a,node distance=30] (coo7aa) {};
\coordinate[below of=coo7b,node distance=45] (coo7bb);
\coordinate[below of=coo7c,node distance=60] (coo7cc);
\coordinate[below of=coo7d,node distance=75] (coo7dd);
\draw[->] (coo5aap) -- (coo5aapp) -- (coo7aa) -- (coo7a);
\draw[->] (coo5bbp) -- (coo5bbpp) -- (coo7bb) -- (coo7b);
\draw[->] (coo5ccp) -- (coo5ccpp) -- (coo7cc) -- (coo7c);
\draw[->] (coo5ddp) -- (coo5ddpp) -- (coo7dd) -- (coo7d);

\draw[->] (coo1) -- node[at start,above,yshift=0mm] {$T_{set}$} (add1);
\draw[->] (add1) -- node[near start,above,yshift=1mm] {$e$}  (Ly);
\draw[->] (coo7aa) -- (coo2) -- node[at end,right] {$-$} (add1);

\coordinate[above of=lhp,node distance=30] (coo8m);
\coordinate[left of=coo8m,node distance=15] (coo8a);
\coordinate[right of=coo8m,node distance=15] (coo8b);
\coordinate[above of=coo8a,node distance=40] (coo8aa);
\coordinate[above of=coo8b,node distance=40] (coo8bb);
\draw[->] (coo8aa) -- node[at start,above,yshift=-0.5mm] {$\dot{Q}_{evap}$} (coo8a);
\draw[->] (coo8bb) -- node[at start,above,xshift=0mm] {$T_{sink}$} (coo8b);
\node[point,above of=coo8a,node distance=30] (coo9a) {};
\node[point,above of=coo8b,node distance=15] (coo9b) {};
\coordinate[left of=coo9a,node distance=100] (coo10a);
\coordinate[left of=coo9b,node distance=100] (coo10b);
\coordinate[above of=Ly,node distance=15] (coo11m);
\coordinate[left of=coo11m,node distance=15] (coo11a);
\coordinate[right of=coo11m,node distance=15] (coo11b);
\draw[->] (coo9a) -- (coo10a) -- (coo11a);
\draw[->] (coo9b) -- (coo10b) -- (coo11b);
\end{tikzpicture}}
	\caption{Block diagram of the nonlinear controller for the LHP control heater}	
	\label{fig:control}
\end{figure}

For the nonlinear state-space system \eqref{eq:ss}, a Lyapunov-based controller is designed. Similar to the control design process in \cite{Gellrich2019}, the controlled variable is the CC temperature, which corresponds to the first state in \eqref{eq:ss}. The manipulated variable $\dot{Q}_{cc}$ as control heater output has only an direct impact on the first state, the CC temperature. The block diagram of the nonlinear feedback control loop is depicted in Fig.~\ref{fig:control}.
\newline
Lyapunov's direct method allows the design of a nonlinear controller, which eliminates the control error
\begin{equation}
\label{eq:error}
e=T_{set}-T_{cc}
\end{equation}
with the setpoint temperature $T_{set}$ being defined in \eqref{eq:Tset}. The chosen candidate $V(e)$ for a positive definite Lyapunov function depends on the control error in \eqref{eq:error}:
\begin{equation}
V(e)=\frac{1}{2}e^2.\label{eq:lyaFct}
\end{equation}
For a decreasing $V(e)$ outside $T_{set}$ along every trajectory $e$, a negative definite derivative $\dot{V}(e)$ of the Lyapunov function is inevitable:
\begin{equation}
\dot{V}(e)=e \cdot\dot{e}=(T_{set}-T_{cc})\cdot (-\dot{T}_{cc}).\label{eq:derivLya}
\end{equation}
The negative definiteness is established when the following equation holds:
\begin{equation}
\dot{T}_{cc}=\lambda\cdot\left(T_{set}-T_{cc}\right)~~~\text{with } \lambda>0.\label{eq:condition}
\end{equation}
The parameter $\lambda$ sets the rate of decrease of the Lyapunov function in \eqref{eq:lyaFct} and is adjusted to the dynamics of \eqref{eq:Tcc}. Inserting \eqref{eq:Tcc} in \eqref{eq:condition} yields:
\begin{equation}
\frac{\dot{m}_l c_p^l\left(T_{cc,in}-T_{cc}\right)+\dot{Q}_{cc}+\dot{Q}_{wick}+A\left(\dot{m}_l-\dot{m}_v\right)}{\left(\rho_{cc}^l  V_{cc}^l+D\rho_{cc}^v\right)c_{p,cc}^l +CD\,\Delta h_{cc}^v-\frac{V_{cc}R}{T_{cc}}B+ACD}=\lambda\cdot e
\end{equation}
with $A$, $B$, $C$, and $D$ defined in \eqref{eq:ABCD}. Thus, the final control law of the control heater results in:
\begin{align}
\dot{Q}_{cc}&=\lambda e\left[\left(\rho_{cc}^l  V_{cc}^l+D\rho_{cc}^v\right)c_{p,cc}^l+CD\,\Delta h_{cc}^v-\frac{V_{cc}R}{T_{cc}}B+ACD\right]\nonumber\\
&-A\left(\dot{m}_l-\dot{m}_v\right)-c_{p}^l\dot{m}_l\left(T_{cc,in}-T_{cc}\right)-\dot{Q}_{wick}.\label{eq:control}
\end{align}

\section{Numerical validation}
\label{sec:valid}
For the validations of the proposed LHP model and the designed controller, the numerical model of \cite{Meinicke2019} is used as LHP~simulation, which was already adapted and used for controller validation in \cite{Gellrich2019}. This LHP~simulation is fitted and validated with measurements from the test bench of the considered LHP and provides a detailed simulation of the LHP temperatures over the entire operating range. The advantage of using an LHP~simulation for validation is the availability of time-variant variables, which cannot be measured in the considered hermetically sealed LHP besides surface temperatures. However, inconsistency in the parameter determination for the introduced state-space model occurs, when the iterative solution method of the LHP~simulation converges to a condenser outlet temperature smaller than the sink temperature $T_{sink}$ due to the discretization of the condenser, which is contrary to the second law of thermodynamics. In this case, the condenser outlet temperature for the parameter determination is set to values slightly higher than the sink temperature to determine the appropriate value of $k_{sc}$ in the OP.

At first, the proposed LHP model (Model~B) is compared to the LHP~simulation (Sim) and to the LHP model (Model~A) of \cite{Gellrich2018b}, whose temperatures are validated with experimental data. Second, the nonlinear Lyapunov-based controller (Controller~B) based on the proposed LHP model is compared in the LHP~simulation to the nonlinear Lyapunov-based controller (Controller~A) of \cite{Gellrich2019}, where the compared LHP model of \cite{Gellrich2018b} is used in the control design.

\subsection{Model validation}\label{sec:modval}

\begin{figure}[htb]
	\centering
	\includegraphics[width=\linewidth]{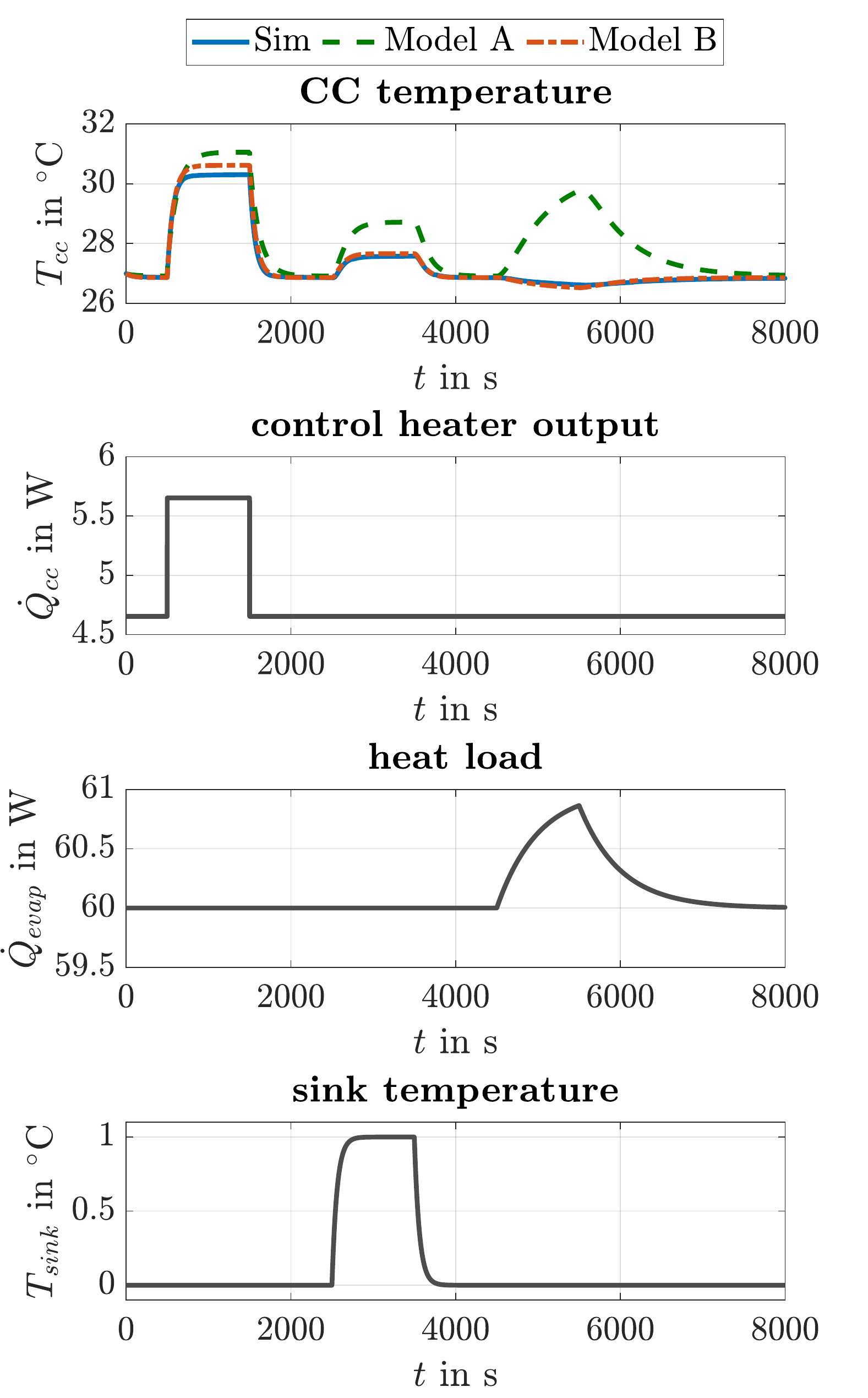} 
	\caption{Comparison of the CC temperatures of Model~A (green), Model~B (red), and the LHP~simulation (blue) to various input changes (the control heater output, the heat load, and the sink temperature)}	
	\label{fig:comp3Models}
\end{figure}

According to the operating range of the considered LHP in Sec.~\ref{sec:fund}, the following OP for the parameter determination (see Sec.~\ref{sec:param}) is chosen from the simulation results:
\begin{table}[htb!]
	\centering
	\small
	\caption{Chosen operating point (OP) from simulated data}
	\label{tb:op}
	$
	\begin{array}{c|c|c}
		\text{inputs} & \text{states} & \text{others}\\\hline	
		\begin{aligned}
			\dot{Q}_{evap}^{op}&=\SI{60.00}{\watt}\\
			T_{sink}^{op}&=\SI{0}{\degreeCelsius}\\
			\dot{Q}_{cc}^{op}&=\SI{4.653}{\watt}\\
			&
		\end{aligned}		 
		&
		\begin{aligned}
			T_{cc}^{op}&=\SI{26.86}{\degreeCelsius}\\
			\eta^{op}&=\SI{0.3268}{\meter}\\
			V_{cc}^{l,op}&=\SI{6.276}{\cubic\centi\meter}\\
			\dot{m}_{l}^{op}&=\SI{50.85}{\milli\gram\per\second}
		\end{aligned}
		&
		\begin{aligned}
		T_{evap}^{op}&=\SI{26.95}{\degreeCelsius}\\
		T_{cond}^{op}&=\SI{26.93}{\degreeCelsius}\\
		T_{cond,out}^{op}&=\SI{0.0001}{\degreeCelsius}\\
		&
		\end{aligned}\rule{0pt}{13mm}\\
		&\\[-3mm]\hline
	\end{array}
	$
\end{table}
\newline
Following the procedure in Sec.~\ref{sec:param}, the four lumped parameters are calculated in the OP as:
\begin{align}
 R_{wick}^{op}&=\SI{0.07720}{\kelvin\per\watt},\\
 k_{2\phi}^{op}&=\SI{1064}{\watt\per\square\meter\per\kelvin},\\
 k_{sc}^{op}&=\SI{312.8}{\watt\per\square\meter\per\kelvin},\\
 k_{ll}^{op}&=\SI{4.804}{\watt\per\square\meter\per\kelvin}.
\end{align}
The pressure losses $\Delta p_{fri}$ in the wick in \eqref{eq:pevap} are determined with the corresponding saturated temperature difference in the chosen OP:
\begin{equation}
 \Delta p_{fri}=\SI{36.71}{\kilo\pascal}.
\end{equation}
The comparison of Model~A and Model~B includes the reaction of the CC temperatures $T_{cc}$ to changes of the input variables, which are the disturbances $\dot{Q}_{evap}$ and $T_{sink}$, and the manipulated variable $\dot{Q}_{cc}$. In Fig.~\ref{fig:comp3Models}, the trajectories of these four variables for both models and the LHP~simulation are presented.
\newline
Under the same influence of the three input variables, the performances of Model~A (dashed green) and B (dashed-dotted red) differ visibly in comparison with the LHP~simulation (solid blue). As expected, Model~A reacts properly to the change of the control heater output with a maximal absolute deviation (MAD) of $\SI{1.01}{\kelvin}$ between $t=0$ and $\SI{2500}{\second}$, but reaches its limitations with changes in the disturbances resulting in absolute deviations up to $\SI{3.17}{\kelvin}$. In addition, the CC temperature of Model~A shows a contrary behavior to the heat load change due to its fixed mass flow rate. However, the CC temperature of Model~B is very similar to the LHP~simulation with both a correct dynamic behavior and a MAD of $\SI{0.34}{\kelvin}$ over the entire simulation time. While the CC dynamics are adapted by directly choosing a suitable lumped thermal capacity of the CC subsystem in Model~A, the CC dynamics in the introduced Model~B depend on the initial liquid volume $V_{cc}^l(t=0)$ in the CC describing yet another impact of the initial phase distribution on the LHP dynamics as on the LHP startup as explained in \cite{Ku2001}.

\begin{figure}[htb]
	\centering
	\includegraphics[width=\linewidth]{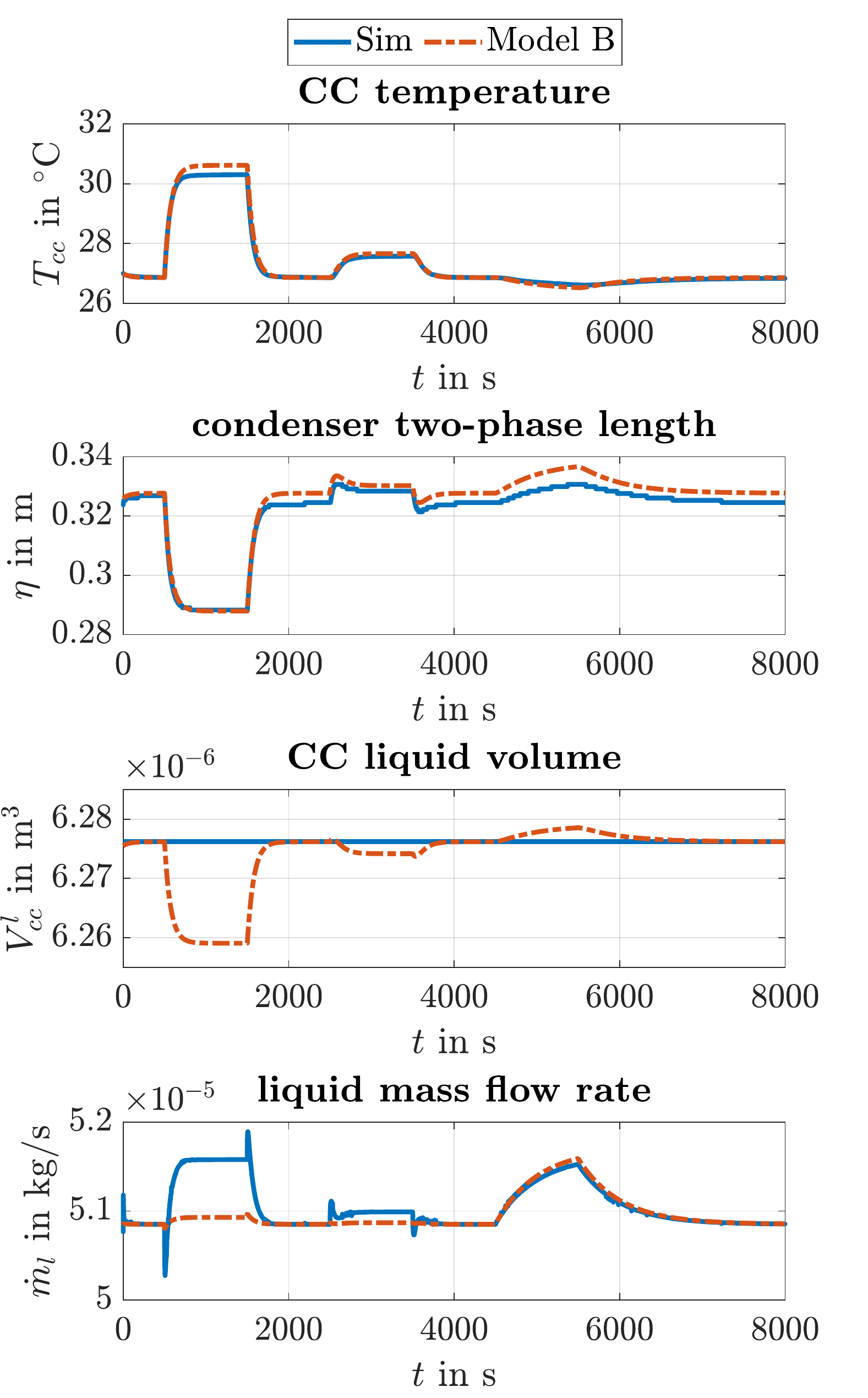} 
	\caption{Comparison of the state variables of Model~B (red) with the corresponding values of the LHP~simulation (blue) to the various input changes in Fig.~\ref{fig:comp3Models}}	
	\label{fig:comp2ModelsStates}
\end{figure}

\begin{figure}[htb]
	\centering
	\includegraphics[width=\linewidth]{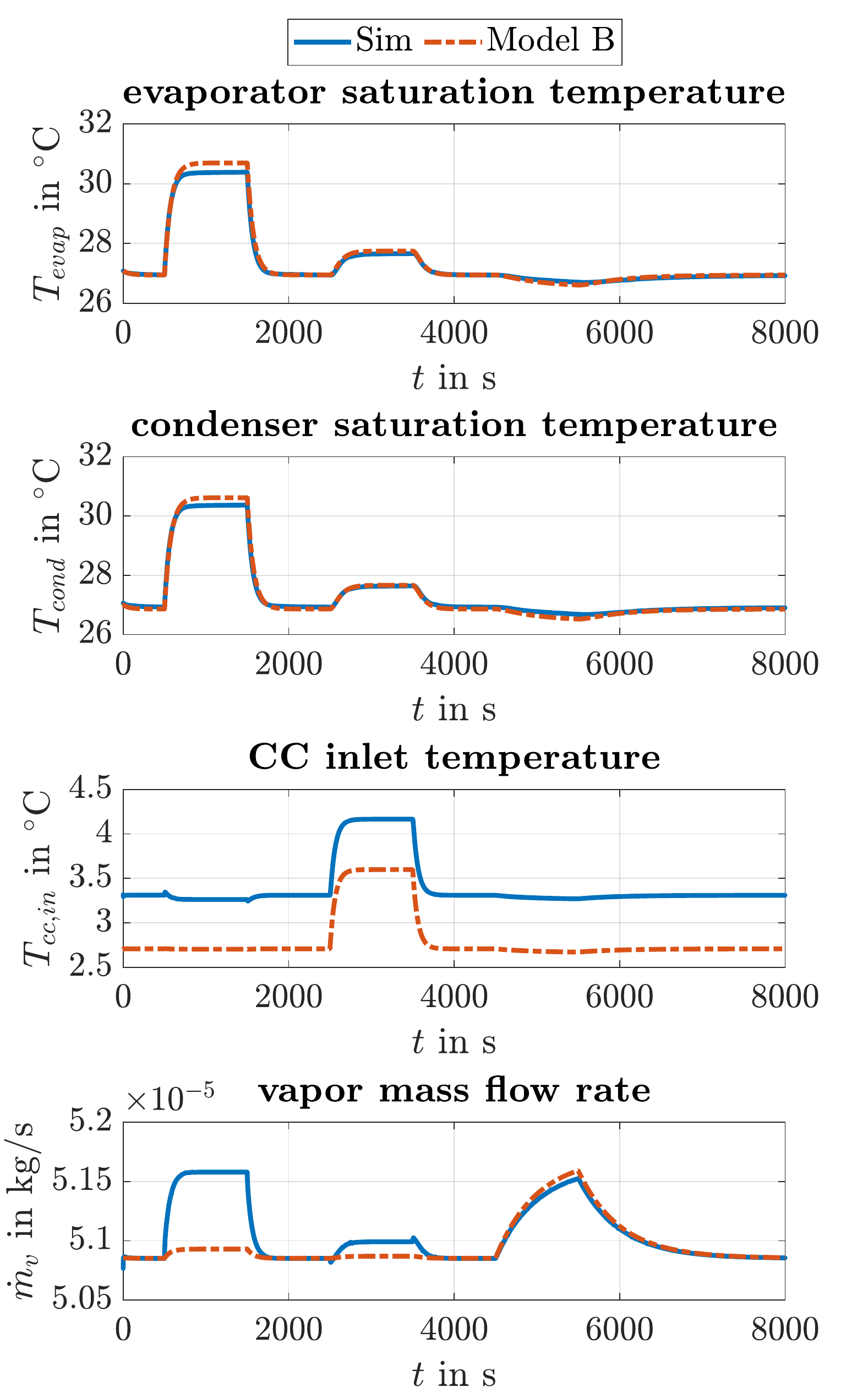} 
	\caption{Comparison of the further variables of Model~B (red) with the corresponding values of the LHP~simulation (blue) to the various input changes in Fig.~\ref{fig:comp3Models}}	
	\label{fig:comp2ModelsMore}
\end{figure}

For a closer look on the differences between Model~B and the LHP~simulation, the state variables of \eqref{eq:ss} are compared to their corresponding values of the LHP~simulation in Fig.~\ref{fig:comp2ModelsStates}.
\newline
As mentioned before, the CC temperature $T_{cc}$ of Model~B (dashed-dotted red) agrees with the CC temperature of the LHP~simulation (solid blue). The lengths $\eta$ of the two-phase region in the condenser of Model~B and of the LHP~simulation are close together with a small MAD of $\SI{0.006}{\meter}$ due to the neglected superheating area in the condenser. Hence, the Assumption~\ref{ass:super} is approved. While the division of the CC fluid volume in the LHP~simulation is fixed, the changes in the CC liquid volume of Model~B are small with a very low impact on the CC temperature. The small deviation between the liquid mass flow rates in Fig.~\ref{fig:comp2ModelsStates} is similar to the deviation in the vapor mass flow rates in Fig.~\ref{fig:comp2ModelsMore}, especially for the first two changes in the input variables.
The reason for these deviations is the heat leak, which is fitted to experimental data in the LHP~simulation as a function of only the heat load and the sink temperature, whereas the heat leak between the evaporator and the CC in Model B is a function of the temperatures of these subsystems. Therefore, the control heater output has also an impact on the heat leak. The different heat leak models cause the offset between the CC inlet temperatures of Model~B and the LHP~simulation in Fig.~\ref{fig:comp2ModelsMore}, because the thermal energy balance in the CC depends on both the CC inlet temperature and the heat leak.

Similar to the CC temperature, the saturation temperatures of the evaporator and the condenser in Fig.~\ref{fig:comp2ModelsMore} are close to their corresponding temperatures of the LHP~simulation. The spikes in the liquid mass flow rate of the LHP~simulation in Fig.~\ref{fig:comp2ModelsStates}, which also appear in the vapor mass flow rate and the CC inlet temperature in Fig.~\ref{fig:comp2ModelsMore}, are much smaller in Model~B. The reason is that the liquid mass flow rate is a function of the control heater output in the LHP~simulation, whereas the control heater input has only a indirect impact on the liquid mass flow rate in Model~B through the CC temperature and its pressure level.

Nevertheless, the dynamics of the LHP variables are precisely modeled in Model~B in order to serve as an adequate control model for the subsequent model-based control design in the next section. Another advantage of Model~B over the LHP~simulation relates to the computational effort. For Model~B, the simulation of the state variables took only $\SI{7}{\second}$, whereas the numerical LHP~simulation terminated after almost $\SI{24}{\minute}$ for a simulation time of $\SI{133}{\minute}$.

\subsection{Control validation}

\begin{figure}[htb]
	\centering
	\includegraphics[width=\linewidth]{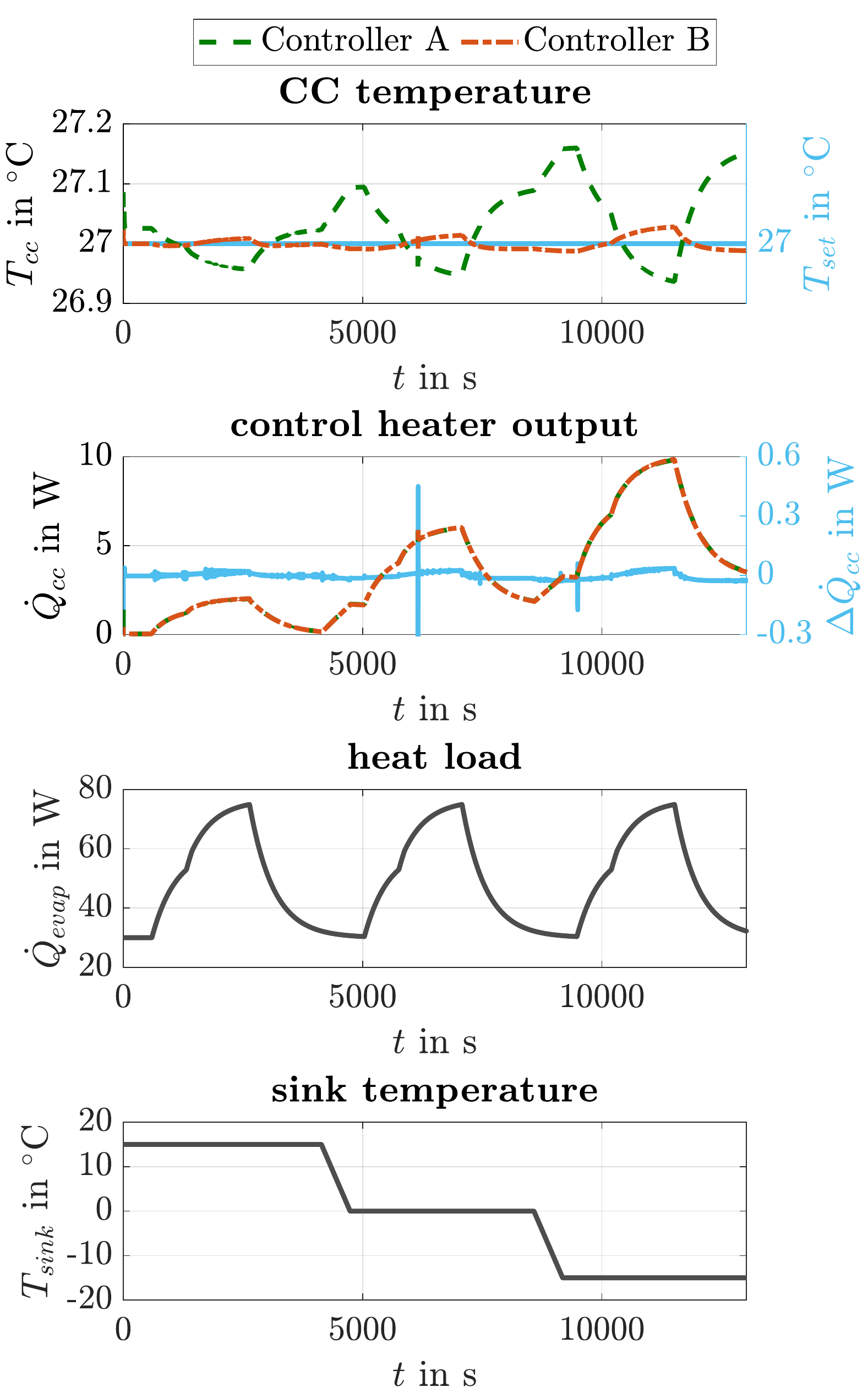} 
	\caption{Comparison of the nonlinear controllers Controller~A (green) and Controller~B (red) on the basis of Model~A and Model~B under varying disturbances (the heat load and the sink temperature)}	
	\label{fig:compControl}
\end{figure}

After the introduced LHP model Model B is validated in Sec.~\ref{sec:modval}, the nonlinear controllers Controller~A and Controller~B based on Model~A and Model~B are tested and evaluated with the LHP~simulation. The control performance of both controllers with the respective disturbance profiles are depicted in Fig.~\ref{fig:compControl}.

The improvement of the controller performance by introducing the fluid's dynamics to the control model are clearly visible. Controller~B with a MAD of $\SI{0.03}{\kelvin}$ between the CC temperature and the setpoint temperature shows an improved disturbance rejection compared to Controller~A with a MAD of $\SI{0.16}{\kelvin}$, which corresponds to an improvement of $\SI{81}{\percent}$. The root mean square error (RMSE) between the CC temperature and the setpoint temperature is improved by $\SI{86}{\percent}$ ($\text{RMSE}_A=\SI{0.07}{\kelvin}$ and $\text{RMSE}_B=\SI{0.01}{\kelvin}$). Although the rates of decrease of both controllers are increased up to the stability limit ($\lambda_A=3$ and $\lambda_B=1$), Controller~A can't reach the robustness of Controller~B. Furthermore, the typical rise of the deviation between the CC temperature and the setpoint temperature with decreasing sink temperature, as mentioned in \cite{Gellrich2018b}, is visible in both control results, since also only the CC temperature is controlled. However, the good agreement of the introduced LHP state-space model with the numerical LHP~simulation mirrors the improved robustness of Controller~B towards Controller~A.

\section{Conclusions}
\label{sec:con}
In this paper, the model-based control design of a nonlinear controller for the control heater of the LHP for temperature control has been presented. The new underlying control model has been deduced from the basic conservation equations of thermodynamic systems as an analytical LHP state-space model. By extending the existing LHP temperature model with the fluid's dynamics, the introduced LHP model includes the dynamic disturbance behavior of the LHP. As a result, the LHP temperatures can be simulated more accurately with the four-dimensional state-space model than with the former state-of-the-art models. In addition, only five unknown parameters must be determined in the operating point. Based on the new LHP state-space model, a nonlinear controller has been designed and compared to the nonlinear controller, which is based on the former LHP temperature model. The control results have shown that the new nonlinear controller improves the disturbance rejection in comparison to the former nonlinear controller.

\appendix

\section{Physical properties of the working fluid}
\label{sec:appendix}
The working fluid of the considered LHP is ammonia. The corresponding values are taken from \cite{Wagner2010}. All temperatures are expressed in \SI{}{\degreeCelsius}.

The density $\rho_l$ of the saturated liquid is approximated with the polynomial
\begin{equation}
\rho_l(T)=-4\cdot 10^{-5}\cdot T^3-0.0027\cdot T^2-1.3522\cdot T+638.57.
\end{equation}
The density $\rho_v$ of the saturated vapor is approximated with the polynomial
\begin{equation}
\rho_v(T)=1\cdot 10^{-5}\cdot T^3-0.0017\cdot T^2+0.1229\cdot T+3.4553.
\end{equation}
The specific heat capacities are:
\begin{equation}
c_p^l(T)=5\cdot 10^{-4}\cdot T^3+3\cdot T^2+5.6\cdot T+4616.5
\end{equation}
for the saturated liquid and
\begin{equation}
c_p^v(T)=0.1\cdot T^2+15.1\cdot T+2680.8
\end{equation}
for the saturated vapor. The heat of evaporation $\Delta h_v$ is given by
\begin{equation}
\Delta h_v(T)=-3\cdot 10^{-2}\cdot T^3-11.5\cdot T^2-3572.3\cdot T+1262300. 
\end{equation}
The dynamic viscosity of the liquid is approximated by the polynomial
\begin{align}
\mu_l(T)&=\left(-2\cdot 10^{-8}\cdot T^5+10^{-6}\cdot T^4-10^{-4}\cdot T^3\right.\nonumber\\
&+0.0151\cdot T^2-1.8665\cdot T+170.1\left.\vphantom{10^{-8}}\right)\cdot 10^{-6}.
\end{align}
The Antoine parameters are
\begin{align}
A_{wf}&=9.394997,\\
B_{wf}&=879.9236,\\
C_{wf}&=-38.15.
\end{align}
The surface tension $\sigma$ of the saturated liquid is (\cite{Kleiber2010})
\begin{equation}
\sigma(T)=0.10175\cdot\left(1-\frac{T+273.15}{405.50}\right)^{1.21703}.
\end{equation}

\bibliographystyle{elsarticle-harv} 
\bibliography{literature/2019_CEP}


%
%
%
\end{document}